\documentclass[11pt,twoside]{article}

\usepackage{epsfig,amsfonts,color}
\usepackage{amsmath}
\usepackage{mathtools}

\usepackage{amssymb, palatino, geometry,url}
\usepackage{amsthm}
\usepackage{algorithmic}
\usepackage{booktabs}
\usepackage{siunitx}
\usepackage[noresetcount,lined,boxed]{algorithm2e} 

\usepackage[colorlinks=true,linkcolor=blue,citecolor=blue,urlcolor=blue]{hyperref}
\usepackage{subcaption}
\usepackage{cleveref}
\usepackage{float}
\usepackage{booktabs}
\usepackage{fancyhdr}
\theoremstyle{definition}

\newcommand{\be}{\begin{equation}}
\newcommand{\ee}{\end{equation}}
\newcommand{\bea}{\begin{eqnarray}}
\newcommand{\eea}{\end{eqnarray}}

\newcommand{\bvec}{\left(\begin{array}{c}}
\newcommand{\evec}{\end{array}\right)}
\newcommand{\bsub}{\begin{subequations}}
\newcommand{\esub}{\end{subequations}}

\usepackage[thinc]{esdiff}

\usepackage{lineno}

\usepackage{outline}

\usepackage{xcolor}
\usepackage{threeparttable}

\usepackage{caption}
\begin{document}

\title{Data Analysis using Riemannian Geometry\\ and Applications to Chemical Engineering}

\author{Alexander Smith${}^{\P}$, Benjamin Laubach${}^{\dagger}$, Ivan Castillo${}^{\dagger}$, and Victor M. Zavala${}^{\P}$\thanks{Corresponding Author: victor.zavala@wisc.edu}\\
{\small ${}^{\P}$Department of Chemical and Biological Engineering}\\
{\small \;University of Wisconsin-Madison, 1415 Engineering Dr, Madison, WI 53706, USA}\\
{\small ${}^{\dagger}$Chemometrics, AI and Statistics}\\
{\small \;The Dow Chemical Company, 332 SH 332 E, Lake Jackson, TX 77566,}}

\date{}
\maketitle

\abstract
We explore the use of tools from Riemannian geometry for the analysis of symmetric positive definite matrices (SPD). An SPD matrix is a versatile data representation that is commonly used in chemical engineering (e.g., covariance/correlation/Hessian matrices and images) and powerful techniques are available for its analysis (e.g., principal component analysis). A key observation that motivates this work is that SPD matrices live on a Riemannian manifold and that implementing techniques that exploit this basic property can yield significant benefits in data-centric tasks such classification and dimensionality reduction. We demonstrate this via a couple of case studies that conduct anomaly detection in the context of process monitoring and image analysis.  

\section{Introduction}

The assumption that data lies in a Euclidean space is pervasive throughout science and engineering and is the basis of diverse data analysis techniques used in these domains. Making this blanket assumption, however, is not always appropriate and can affect the accuracy/interpretability of such techniques or even break fundamental physical laws. 
\\

Recognizing that data can live in spaces that are governed by \emph{non-Euclidean} geometry is critical to appropriately representing, manipulating, and analyzing certain data objects. A simple example of this arises when computing travel paths between a couple of points that are located on the surface of the Earth; when computing the distance between such points, the \emph{elliptic} geometry of the Earth surface must be taken into consideration. If Euclidean geometry is assumed, travel paths between antipodal locations (e.g., United States and China) can require traversing the Earth (not physically-realizable paths). 
\\

Elliptic geometry is a non-Euclidean geometry in which one of the postulates of Eucliean geometry (sum of interior angles of a triangle is equal to 180${}^o$) no longer holds, because of the presence of positive curvature in a surface  \cite{chavel2006riemannian}. Elliptic geometry was first proposed by Bernhard Riemann in the 19th century, and he further developed these ideas into the field that is now known as Riemannian geometry \cite{riemann2016hypotheses}. 
\\

An important example in which assuming Euclidean geometry can lead to spurious results is in the analysis of \emph{symmetric positive definite} (SPD) matrices (e.g., covariance/correlation matrices). SPD matrices lie on a high-dimensional space which is governed by Riemannian geometry (known as a Riemannian manifold) \cite{lee2006riemannian}.  Standard techniques for the analysis of SPD matrices (e.g., PCA or basic matrix norms) do not take this property into consideration and can lead to misleading results.  For instance, computing the distance between SPD matrices in Euclidean geometry (e.g., via the Frobenius norm) ignores the fact that such matrices live on a Riemannian manifold, and this can yield misleading results  \cite{fletcher2004principal,arsigny2007geometric,sommer2010manifold}.  Specifically, the so-called  \emph{swelling effect} can occur when applying operations in Euclidean geometry to SPD matrices \cite{arsigny2007geometric}. This effect introduces spurious results by inflating the determinants of SPD matrices and can also distort the results of commonly used methods  \cite{sommer2010manifold}. Computing interpolations and averages of SPD matrices, which is key in understanding physical systems (e.g. Brownian motion), can also break physical conservation laws if performed under Euclidean geometry \cite{pennec2020manifold}. 
\\

In this work, we focus our attention on the use of techniques from Riemannian geometry for the analysis of data objects that can be represented as SPD matrices. An SPD matrix is a simple but versatile data representation that is widely used in multivariate analysis techniques such as PCA  \cite{reinsel2003elements,wasserman2004all,yao2015sample}. SPD  representations are also used in process control, monitoring, and anomaly detection  \cite{feng2014kalman,iruthayarajan2010covariance,mansouri2016statistical,wise1996process,russell2000fault,smith2021euler}, in the study of functional brain networks \cite{qiu2015manifold, sporns2002network,varoquaux2010detection,goni2014resting,smith2021euler}, in object detection \cite{tuzel2006region,porikli2006achieving, xu2015anomaly}, in biomedical image analysis \cite{pennec2020manifold, portela2014semi}, in the analysis of Laplacian matrices in graph theory, and in the analysis of Hessian matrices in optimization \cite{nocedal2006numerical}. In applications such as image analysis, an SPD representation can be obtained by applying transformations to a raw data object (e.g., smoothing via kernel functions and/or combination of image features). The defining feature of an SPD matrix is that all its eigenvalues are real and positive. In the context of optimization, it is well-known that an SPD Hessian matrix defines positive curvature of a multi-dimensional quadratic function and is key in defining the geometry of objective and constraint functions (e.g., convex or non-convex). In the context of statistics, it is well-known that an SPD covariance matrix defines a multi-dimensional ellipsoid (a surface with positive curvature) and that the level sets of a multivariate Gaussian probability density function are ellipsoids. 
\\

This paper provides a practical introduction to the Riemannian geometry of SPD matrices and demonstrates applications of interest to the process systems engineering community. Specifically, we illustrate the benefits of exploiting the Riemannian geometry of SPD matrices and discuss how these tools can be incorporated into common dimensionality reduction and classification workflows. We also provide example applications of interest in science and engineering. The first application focuses on analysis of covariance matrices derived from multivariate time series, which is a common task in process monitoring. Our application focuses on the so-called Tennessee Eastman Process (TEP) \cite{downs1993plant}. The TEP is a process where anomalies/faults are systematically introduced which shifts the relationships between the measured variables. Covariance matrices encode these changing relationships and are then used to predict what type of anomaly the process is experiencing. The second application is in defect/anomaly detection of textiles taken from the MVTEC AD dataset \cite{bergmann2019mvtec}. Here, grayscale images of textiles are represented as covariance matrices by characterizing the relationships between the original image and multiple transformations of the image (e.g., smoothing with different kernels). The image transformations emphasize different features of the original image (e.g., edges and fibers). Subsequently, when an anomaly/defect is introduced (e.g., a cut or discoloration of the textile) these relationships will change, impacting the covariance matrix and allowing us to detect anomalies. This type of analysis can also be applied to other relevant image/field datasets such as those arising in microscopy and flow cytometry \cite{smith2021euler,smith2021topological}. All data and scripts needed to conduct such analyses is provided as open-source code in \url{https://github.com/zavalab/ML/tree/master/RiemannianSPD}. With this, we aim to provide a concise and easy introduction to non-experts to the field of Riemannian geometry. 

\section{Riemannian Manifolds}

The key observation driving this work is that SPD matrices lie on a Riemannian manifold; we thus begin our discussion by characterizing such manifolds. In this section, we aim to provide an intuitive understanding of manifolds and equip the reader with knowledge of their key properties. 
\\

An $n$-dimensional manifold is a topological space (a space where closeness and connectedness are defined but not directly measurable) that {\em locally} resembles an $n$-dimensional Euclidean space. Specifically, an $n$-dimensional manifold $M$ is a set where every point $p \in M$ has an open neighborhood $U \subset M$ (known as a \emph{chart}) that can be mapped to an open set of $n$-dimensional Euclidean space $V \subset \mathbb{R}^n$ via a one-to-one, onto, and continuous mapping $f: U \rightarrow V$ (i.e., a homeomorphism). The set of charts whose union covers the manifold is known as an \emph{atlas} $\bigcup_{i=1}^n U_i = M$. The chart/atlas nomenclature is derived from navigation along the surface of the Earth (a $2$D manifold); here, charts are flat (Euclidean) maps of the Earth that are collected in an atlas. 
\\

\begin{figure}[!h]
  \centering
    \begin{subfigure}{.31\textwidth}
      \centering
      \includegraphics[width=.9\linewidth]{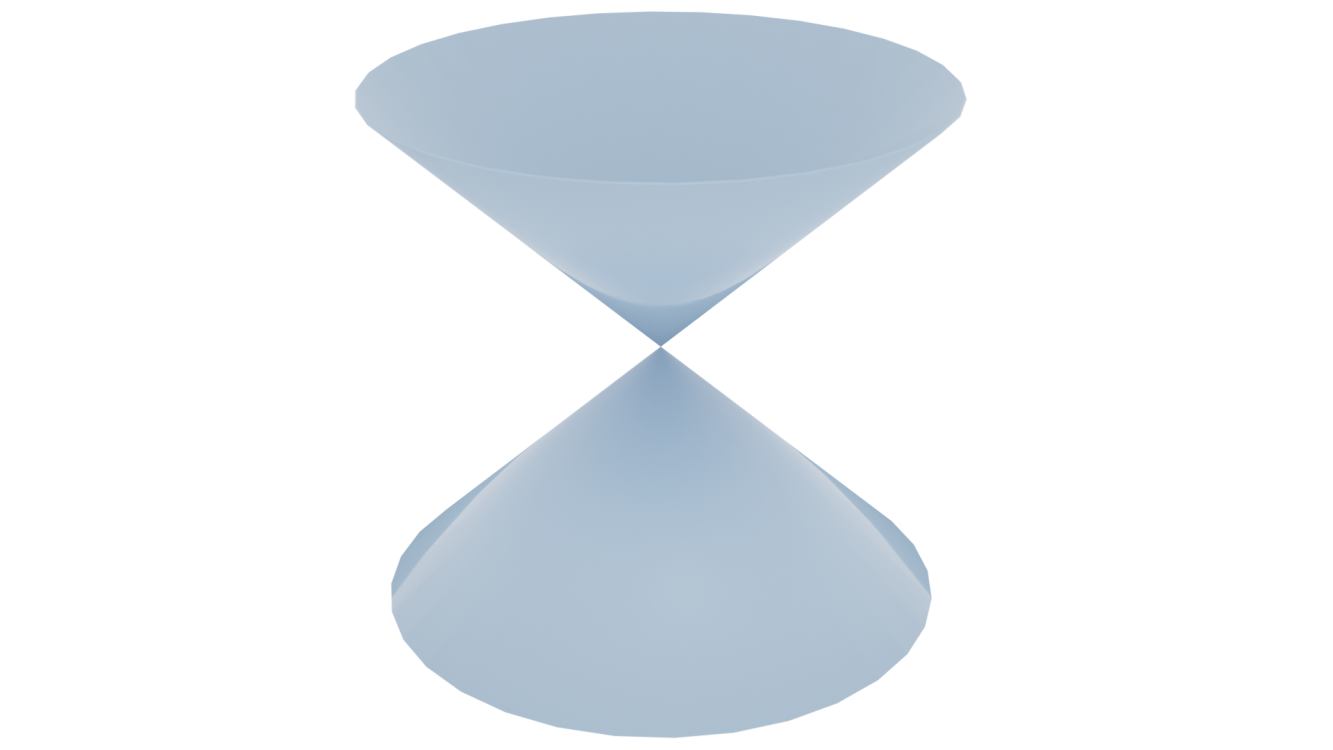}  
      \caption{Non-Manifold}
      \label{fig:sub-first}
    \end{subfigure}
    \begin{subfigure}{.31\textwidth}
      \centering
      \includegraphics[width=.9\linewidth]{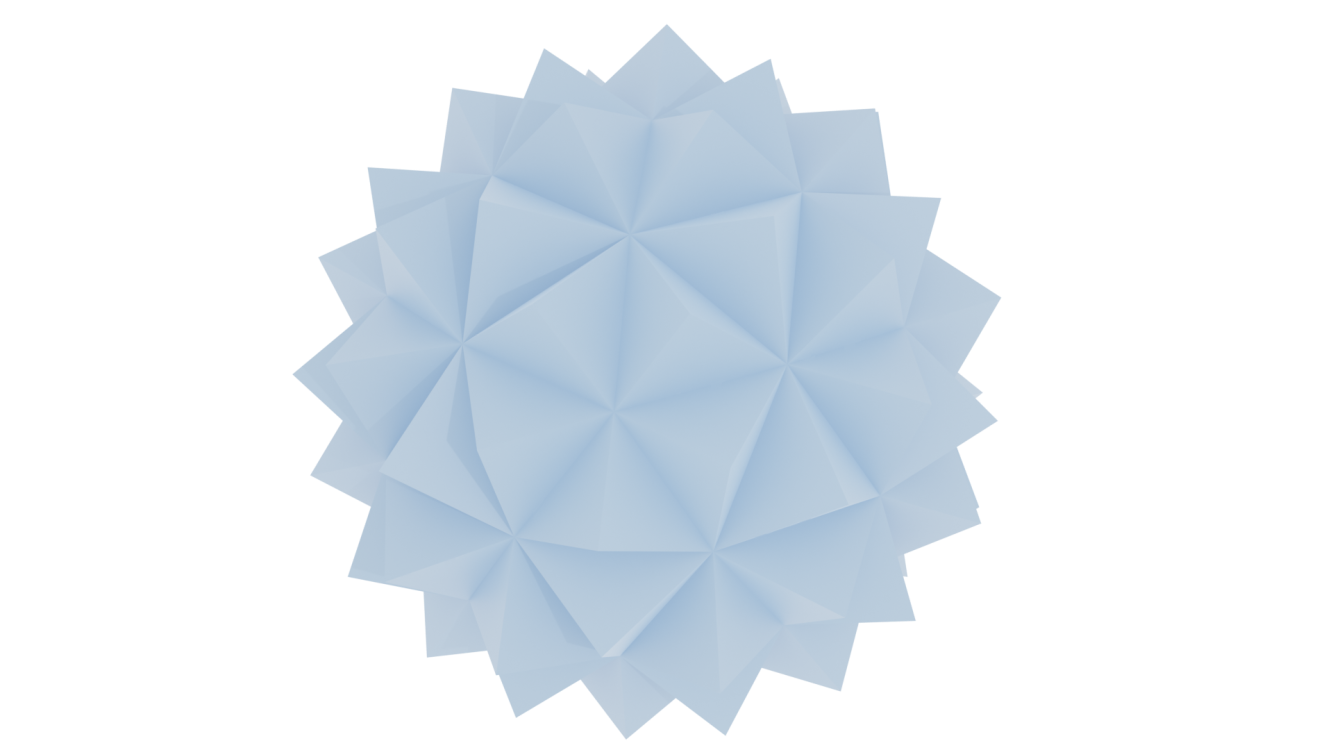}  
      \caption{Non-Differentiable Manifold}
      \label{fig:sub-second}
    \end{subfigure} 
    \begin{subfigure}{.31\textwidth}
      \centering
      \includegraphics[width=.9\linewidth]{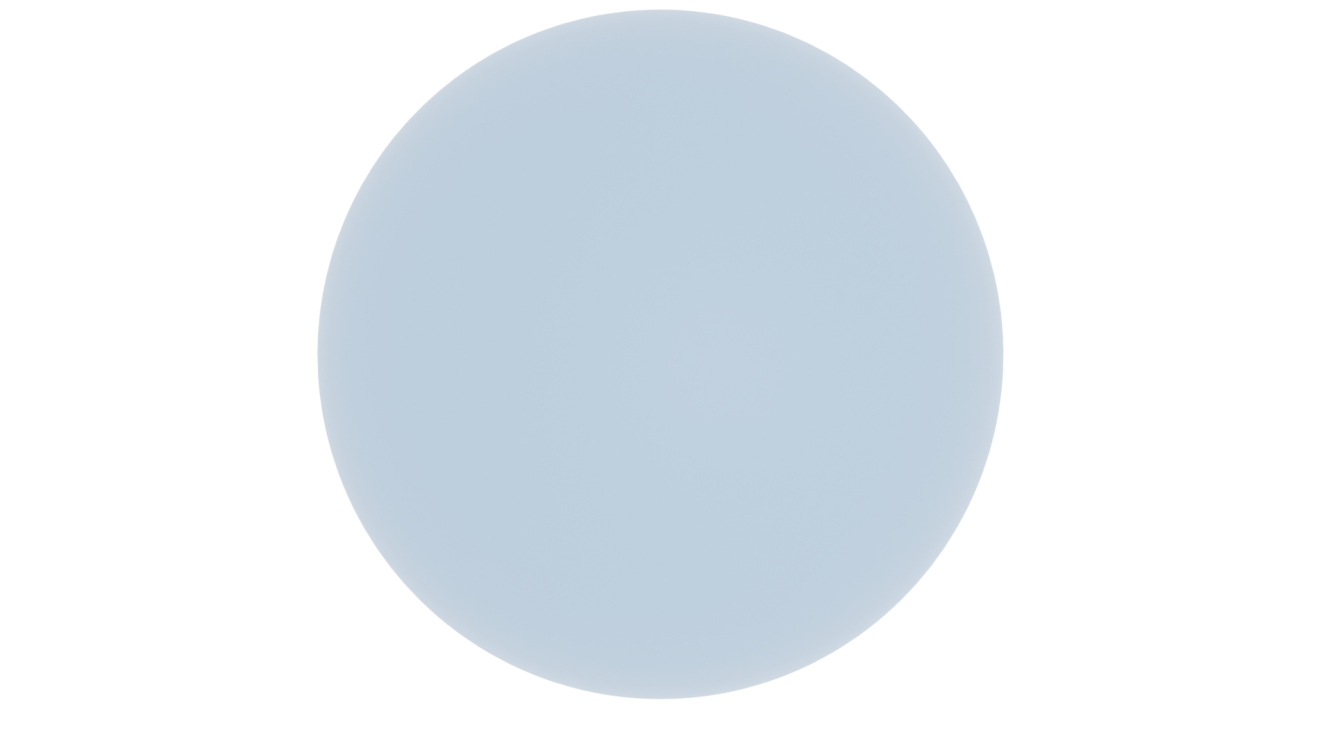}  
      \caption{Differentiable Manifold}
      \label{fig:sub-second}
    \end{subfigure} 
\caption{(a) Space composed of a couple of cones intersecting at a single point. This is a non-manifold space because any neighborhood formed around the intersecting point is not homeomorphic to 2D Euclidean space (the neighborhood is a smaller version of the two intersecting cones). (b) Represents a 2D manifold (all points and associated neighborhoods can be mapped to 2D  Euclidean space) but is not a differentiable manifold because of the cusps occurring at the edges of the manifold (differential is not defined everywhere). (c) A smooth sphere is a 2D manifold that is also differentiable (curves on the surface can be differentiated everywhere).}
\label{fig:3D_Reg}
\end{figure}

In Figure \ref{fig:3D_Reg} we illustrate multiple topological spaces embedded in a 3D Euclidean space. The spiked (b) and smooth (c) hollow spheres are examples of $2$D manifolds. The space in (a) represents a couple of cones that intersect at a single point and is not manifold; this is because a neighborhood drawn around the intersecting point will look like a smaller version of the intersecting cones, which cannot be mapped to $2$D  Euclidean space through a homeomorphism. 

Manifolds can also be endowed with geometric structure; for our analysis, we are particularly interested in whether or not a given manifold is \emph{differentiable}. In simple terms, a differentiable manifold is a manifold for which calculus (e.g., computing derivatives and integrals) can be performed on the charts that make up the manifold atlas \cite{lang2012fundamentals}. This also means that \emph{curves} on the surface of the manifold can be analyzed from a geometric perspective using calculus. A curve is defined as a continuous function $\gamma: [a,b] \rightarrow M$ mapping the interval $[a,b] \in \mathbb{R}$ to the manifold $M$. We will not cover the specific mathematical requirements that make a manifold differentiable in general, but refer interested readers to the following reference for details \cite{lang2012fundamentals}. Examples of a differentiable and non-differentiable manifold are shown in Figure \ref{fig:3D_Reg}. We can see that the spiked sphere in (b) has multiple cusps where a derivative cannot be defined for a curve, making it non-differentiable. For the smooth sphere in (c), a derivative can be taken anywhere, allowing for more complex operations/transformations to be performed on the manifold \cite{pennec2006intrinsic}.
\\

\begin{figure}[!h]
  \centering
    \begin{subfigure}{.49\textwidth}
      \centering
      \includegraphics[width=.9\linewidth]{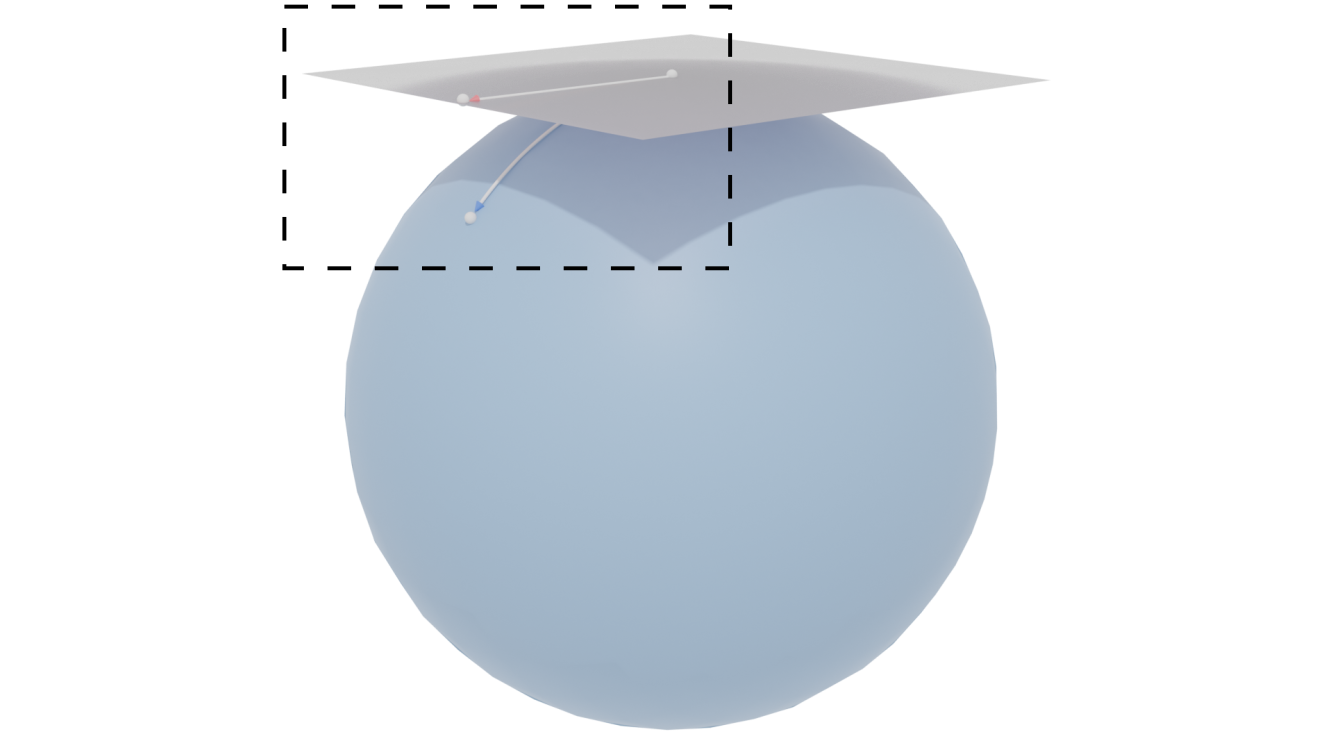}  
      \caption{}
      \label{fig:sub-first}
    \end{subfigure}
    \begin{subfigure}{.49\textwidth}
      \centering
      \includegraphics[width=.9\linewidth]{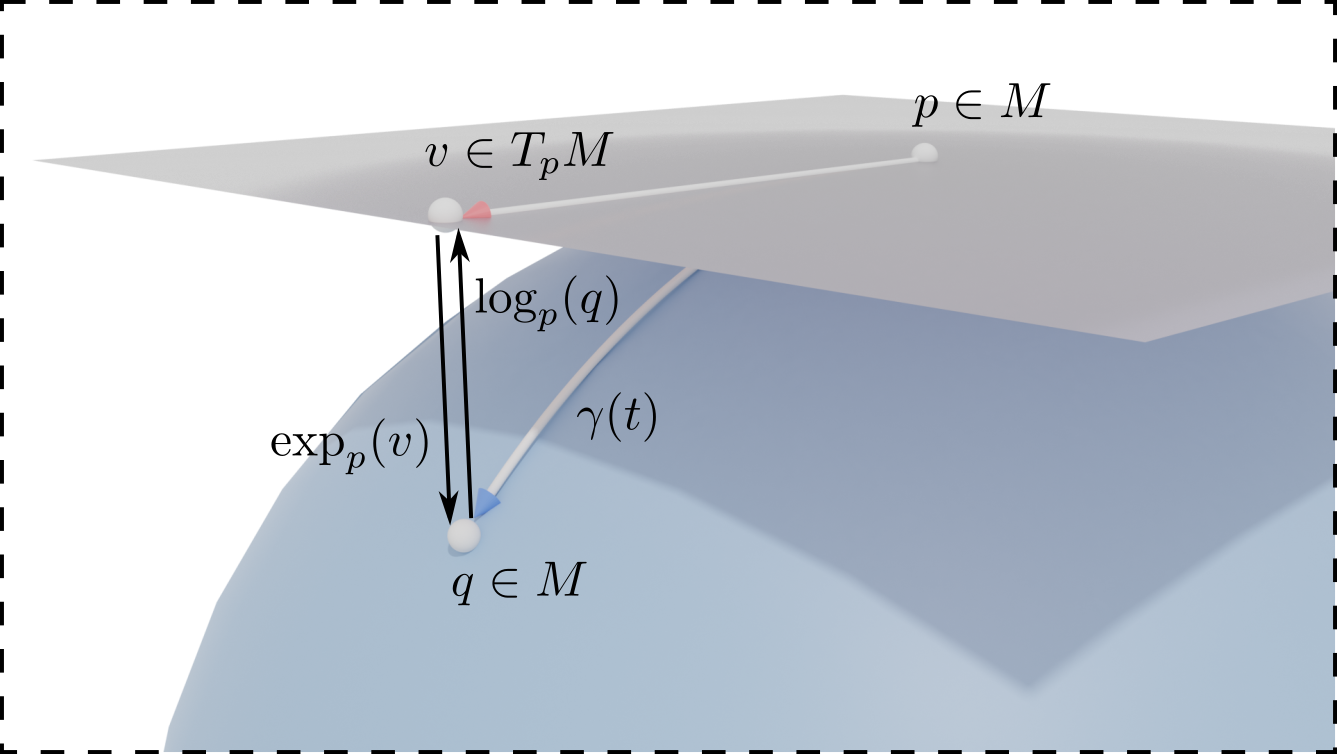}  
      \caption{}
      \label{fig:sub-second}
    \end{subfigure} 
  \caption{(a) Illustration of a Riemannian manifold ($M$) and the associated tangent space at a point $p \in M$. (b) Illustration of a geodesic $\gamma(t)$ constructed between two points $p,q \in M$, along with the associated tangent space vector $v \in T_pM$. The exponential map ($\text{exp}_p(v) : v \rightarrow q$) and the logarithmic map ($\text{log}_p(q) : q \rightarrow v$) are also shown.}
  \label{fig:Rieman_Examp}
\end{figure}

We now restrict our attention to a special class of manifolds known as \emph{Riemannian manifolds}. A Riemannian manifold is a differentiable manifold equipped with a defined \emph{tangent space} at each point in the manifold $p \in M$, denoted as $T_pM$ \cite{lee2006riemannian}. The tangent space is the set of tangent vectors of all curves passing through point $p \in M$. The tangent space $T_pM$ is a vector (linear) space that is of the same dimension as the manifold itself. An illustration of a tangent space is shown in Figure \ref{fig:Rieman_Examp} for a smooth sphere (a $2$D Riemannian manifold). For a Riemannian manifold, the tangent space $T_pM$ is equipped with an inner product $g_p : T_pM \times T_pM \rightarrow \mathbb{R}$, along with a norm metric $|\cdot|_p : T_pM \rightarrow \mathbb{R}$ defined by $|v|_p = \sqrt{g_p(v,v)}$ for any vector $v \in T_pM$. These properties allow us to define the \emph{length} of a curve on the manifold surface. A differentiable curve $\gamma: [a,b] \rightarrow M$ assigns to each $t \in (a,b)$ a tangent vector $\gamma'(t) \in T_{\gamma(t)}M$; thus, to obtain the length of the curve $L(\gamma)$, we integrate the norm of the tangent vectors along the curve (i.e., arc length):

\begin{equation}
L(\gamma) := \int_{a}^{b} |\gamma'(t)|_{\gamma(t)} \ dt
\end{equation}

In our analysis, we are primarily interested in measuring the \emph{shortest} curve between a couple points on the manifold (known as a \emph{geodesic}). Given a couple of points on a Riemannian manifold $p,q \in M$ and the set of all curves $\gamma : [a,b] \rightarrow M$ such that $\gamma(a) = p$ and $\gamma(b) = q$, the geodesic $\bar{\gamma}$ is the curve with the shortest total length $L(\bar{\gamma})$:

\begin{equation}
L(\bar{\gamma}) := \text{inf}\{L(\gamma) \ | \ \gamma: [a,b] \rightarrow M, \text{with} \ \gamma(a) = p, \gamma(b) = q \}
\end{equation}

An illustration of the geodesic between a couple of points on the smooth sphere is shown in Figure \ref{fig:Rieman_Examp}. Geodesics are a powerful tool for quantifying the relationship between points on a manifold surface and can be used to compute summarizing statistics for points on the surface (such as means and variances) \cite{pennec2006intrinsic}. 

There are direct relationships between the tangent space of a Riemannian manifold and geodesics, such as the \emph{exponential map} and the \emph{logarithmic map}. For a tangent vector $v \in T_pM$ constructed at point $p \in M$, there exists a unique geodesic $\gamma: [0,1] \rightarrow M$ such that $\gamma(0) = p$ and $\gamma'(0) = v$. The vector $v \in T_pM$ is mapped to the endpoint of the geodesic $\gamma(1) \in M$ through the exponential map (see Figure \ref{fig:Rieman_Examp}):

\begin{align}
\text{exp}_p(v) = \gamma(1). 
\end{align}

The inverse of the exponential map is the logarithmic map, which maps the point $\gamma(1)$ in the neighborhood of $p \in M$ to a vector in the tangent space  $v \in T_pM$ (see Figure \ref{fig:Rieman_Examp}):

\begin{align}
\text{log}_p(\gamma(1)) = v
\end{align}

These functions provide a mapping from the surface of the manifold to the tangent space of a given point. The tangent space, which is a (linear) vector space, can be analyzed using standard techniques designed for Euclidean spaces (e.g., classification, regression, and dimensionality reduction) while properly capturing the relationships between points defined by the geometry of the manifold (e.g., geodesics). For instance, we can use these constructs to analyze and operate on the space of SPD matrices, as we discuss next. 

\section{Riemannian Geometry of SPD Matrices}

This section will provide the mathematical background needed to understand the geometry of the space defined by SPD matrices. We first introduce the needed notation and define the properties of different matrix spaces that are used in our analysis. We then demonstrate how the properties of these matrix spaces can be used to quantify the geometry and structure of the space defined by SPD matrices. The main message is that SPD matrices lie on a Riemannian manifold and that important computations (e.g., matrix operations, summarizing statistics, classification, regression, and dimensionality reduction) can be performed by respecting the geometry of this manifold by conducting these on the tangent space;  respecting such properties can lead to important improvements in efficiency and interpretability. 

\subsection{Matrix Spaces, Properties, and Notation}

We define spaces and properties of matrices that reflect the structure of SPD matrices. We denote $\mathcal{S}(n) := \{\mathbf{S} \in \mathcal{M}(n), \mathbf{S} = \mathbf{S}^T\}$ as the set of symmetric $n\times n$ matrices in the space of square, real matrices $\mathcal{M}(n)$ and the set $\mathcal{P}(n) := \{ \mathbf{P} \in \mathcal{S}(n), u^T\mathbf{P} u > 0, \forall \ u \in \mathbb{R}^n\}$ as the set of all $n \times n$ SPD matrices. 
\\

We also define the exponential and logarithmic mappings used in our analysis. The matrix exponential $\text{exp}(\mathbf{P})$, where $\mathbf{P} \in \mathcal{S}(n)$, is defined as:

\begin{equation}
\text{exp}(\mathbf{P}) := \mathbf{U} \ \text{diag}(\text{exp}(\lambda_1), ..., \text{exp}(\lambda_n)) \ \mathbf{U}^T
\end{equation}

\noindent where $\mathbf{U}$ represents the matrix of eigenvectors of $\mathbf{P}$ and $\lambda_1 > ... > \lambda_n$ represent the eigenvalues of $\mathbf{P}$ (also denoted as $\lambda_i(\mathbf{P})$). We also define the inverse operation; the matrix logarithm $\text{log}(\mathbf{P})$ as:

\begin{equation}
\text{log}(\mathbf{P}) := \mathbf{U} \ \text{diag}(\text{log}(\lambda_1), ..., \text{log}(\lambda_n)) \ \mathbf{U}^T.
\end{equation}

\noindent The following properties should also be considered in the analysis \cite{moakher2006symmetric}:

\begin{itemize}
	\item{$\forall \ \mathbf{P} \in \mathcal{P}(n)$ we have that $\text{det}(\mathbf{P}) > 0$}
	\item{$\forall \ \mathbf{P} \in \mathcal{P}(n)$ we have that $\mathbf{P}^{-1} \in \mathcal{P}(n)$} 
	\item{$\forall \ \mathbf{P} \in \mathcal{P}(n)$ we have that $ \text{log}(\mathbf{P}) \in \mathcal{S}(n)$}
	\item{$\forall \ \mathbf{S} \in \mathcal{S}(n)$ we have that $\text{exp}(\mathbf{S}) \in \mathcal{P}(n)$}
\end{itemize}

\noindent We also define the Frobenius inner product for matrices $\mathbf{A}$ and $\mathbf{B}$ as:

\begin{equation}
\langle \mathbf{A}, \mathbf{B} \rangle_F := \text{Tr}(\mathbf{A}^T\mathbf{B}),
\end{equation}

\noindent where Tr($\cdot$) represents the matrix trace operator. The Frobenius norm for a matrix $\mathbf{A}$ is:

\begin{align}
||\mathbf{A}||_F = \sqrt{\text{Tr}(\mathbf{A}^T\mathbf{A})} = \sqrt{\sum_{i=1}^n \lambda_i(\mathbf{A})}.
\end{align}

\subsection{Manifold of SPD Matrices}

To understand the Riemannian geometry of the space of SPD Matrices ($\mathcal{P}(n)$), we first need to construct a Riemannian {\em metric}, which will allow us to computed distances and other relationships between points in a manifold. The metric we consider for Riemannian manifolds is known as the {\em Affine Invariant Riemannian Metric (AIRM)}; a detailed derivation of the metric is found in the work of Bhatia \cite{bhatia2009positive}. We aim to provide an intuitive understanding of this metric, and its meaning. 
\\

Given SPD matrices $\mathbf{A} \in \mathcal{P}(n)$ and $\mathbf{B} \in \mathcal{P}(n)$, we construct a geodesic $\gamma(t): [0,1] \rightarrow \mathcal{P}(n)$ parameterized as:

\begin{align}
\gamma(t) = \text{exp} \big( (1-t)\cdot \text{log}(\mathbf{A}) + t\cdot \text{log}(\mathbf{B}) \big)
\end{align}

\noindent where we have $\text{log}$ maps $\mathcal{P}(n) \rightarrow \mathcal{S}(n)$ and  $\text{exp}$ maps $\mathcal{S}(n) \rightarrow \mathcal{P}(n)$. Here, we are using the $n \times n$ identity matrix $\mathbf{I} \in \mathcal{P}(n)$ as our tangent space basis. Informally, we are mapping our matrices from our SPD manifold $\mathcal{P}(n)$ to the tangent vector space $\mathcal{S}(n)$ via the logarithmic map, constructing a line between these points in the tangent space, and then projecting the constructed line back to the manifold via the exponential map. This is guaranteed to be the shortest length path between the points in $\mathcal{P}(n)$. The corresponding geodesic distance between matrices $\mathbf{A}$ and $\mathbf{B}$ is given by:

\begin{align}
d_g(\mathbf{A},\mathbf{B}) := ||\text{log}(\mathbf{A}) - \text{log}(\mathbf{B})||_F.
\end{align}

Here, note that we are simply projecting the matrices from the Riemannian manifold to the tangent vector space prior to measuring their distance using the Frobenius norm  \cite{bhatia2009positive}. However, an important consideration must be made when using the $n \times n$ identity matrix $\mathbf{I}$ as the tangent space basis. In many applications, the data may lie within a particular neighborhood that is far from $\mathbf{I}$ on the manifold. Thus, projections to the tangent space $T_IM$ can result in distortions of the data \cite{pennec2006intrinsic}. Intuitively, one can think of this as similar to an analysis of a projection of the Earth surface onto a plane tangent to the 0' latitude and 0' longitude point (as many maps are represented). In this projection, landmasses near the edge of the projection are highly distorted, whereas points near (0',0') have almost no distortion. Thus, our aim is to identify a distance with respect to a tangent space defined by the matrices of concern. This distance can be constructed using a critical property of this Riemannian manifold and metric: \emph{congruence invariance}. 
\\

Congruence invariance states that, for any $n \times n$ invertible matrix $\mathbf{X}$ and matrices $\mathbf{A},\mathbf{B} \in \mathcal{P}(n)$:

\begin{align}
d_g(\mathbf{X}^T \mathbf{A} \mathbf{X}, \mathbf{X}^T \mathbf{B} \mathbf{X}) = d_g(\mathbf{A},\mathbf{B})
\end{align}

We thus have that linear transformations of the given matrices do not impact the geodesic distance on the manifold. This property allows us to redefine the  geodesic distance as:
\begin{subequations}
\begin{align}
d_g(\mathbf{A},\mathbf{B}) & = d_g(\mathbf{I}, \mathbf{A}^{-1/2}\mathbf{B}\mathbf{A}^{-1/2}) \\
& = ||\text{log}(\mathbf{I}) - \text{log}(\mathbf{A}^{-1/2}\mathbf{B}\mathbf{A}^{-1/2})||_F \\
& = ||\text{log}(\mathbf{A}^{-1/2}\mathbf{B}\mathbf{A}^{-1/2})||_F \\
& = \bigg(\sum_{i=1}^n \text{log}^2 \lambda_i(\mathbf{A}^{-1}\mathbf{B})\bigg)^{1/2}
\end{align}
\end{subequations}
\noindent where $\mathbf{I}$ is the $n \times n$ identity matrix, $\mathbf{A} = \mathbf{A}^{1/2}\mathbf{A}^{1/2}$, and $\lambda_i(\mathbf{A}^{-1}\mathbf{B})$ are the eigenvalues of $\mathbf{A}^{-1}\mathbf{B}$. 
\\

We can simplify the geodesic distance further; we have that the geodesic between matrices $\mathbf{I}$ and $\mathbf{A}^{-1/2}\mathbf{B}\mathbf{A}^{-1/2}$ is:

\begin{align}
\gamma_0(t) = \text{exp} \big(\text{log}(\mathbf{A}^{-1/2}\mathbf{B}\mathbf{A}^{-1/2})t \big) = (\mathbf{A}^{-1/2}\mathbf{B}\mathbf{A}^{-1/2})^t
\end{align}

We can then leverage congruence invariance to shift this geodesic to our matrices of interest as:

\begin{align}
\gamma(t) = \mathbf{A}^{1/2}\big(\gamma_0(t)\big)\mathbf{A}^{1/2} = \mathbf{A}^{1/2}(\mathbf{A}^{-1/2}\mathbf{B}\mathbf{A}^{-1/2})^t \mathbf{A}^{1/2}
\end{align}

where $\gamma(0) = \mathbf{A}$ and $\gamma(1) = \mathbf{B}$, which provides a geodesic that is independent of the tangent space basis. Essentially, we are leveraging congruence invariance to translate our points to a neighborhood of $\mathbf{I}$, which allows us to compute distances with minimal distortion, and then translate the points back through these linear transformations. We can apply this same logic to the exponential map (and the logarithmic map); for matrices $\mathbf{A}, \mathbf{B} \in \mathcal{P}(n)$ and $\mathbf{T}_B \in T_AM$ where $T_AM \subset \mathcal{S}(n)$:

\begin{align}
\mathbf{B} = \text{exp}_{A} (\mathbf{T_B}) = \mathbf{A}^{1/2} \text{exp} (\mathbf{A}^{-1/2}\mathbf{T_B}\mathbf{A}^{-1/2}) \mathbf{A}^{1/2}
\end{align}

\begin{align}
\mathbf{T_B} = \text{log}_{A} (\mathbf{B}) = \mathbf{A}^{1/2} \text{log} (\mathbf{A}^{-1/2}\mathbf{B}\mathbf{A}^{-1/2}) \mathbf{A}^{1/2}
\end{align}

Here, we are mapping the matrix $\mathbf{B} \in \mathcal{P}(n)$ to the tangent space vector $\mathbf{T}_B \in T_AM$ which is centered at $\mathbf{A}\in \mathcal{P}(n)$ via $\text{log}_A(\mathbf{B})$ and inversely through $\text{exp}_A(\mathbf{B})$. Thus, with these newly defined geodesics and mappings, we are able to compute relationships between SPD matrices with minimal distortions in the tangent space. However, when given a large dataset with multiple SPD matrices, the choice of a tangent space basis may not be immediately clear. In this case, the geometric mean of the matrices on the manifold is typically identified and used as reference point. 

\subsection{SPD Matrix Means and Tangent Spaces}

\begin{figure}[!h]
  \centering
    \begin{subfigure}{.49\textwidth}
      \centering
      \includegraphics[width=.9\linewidth]{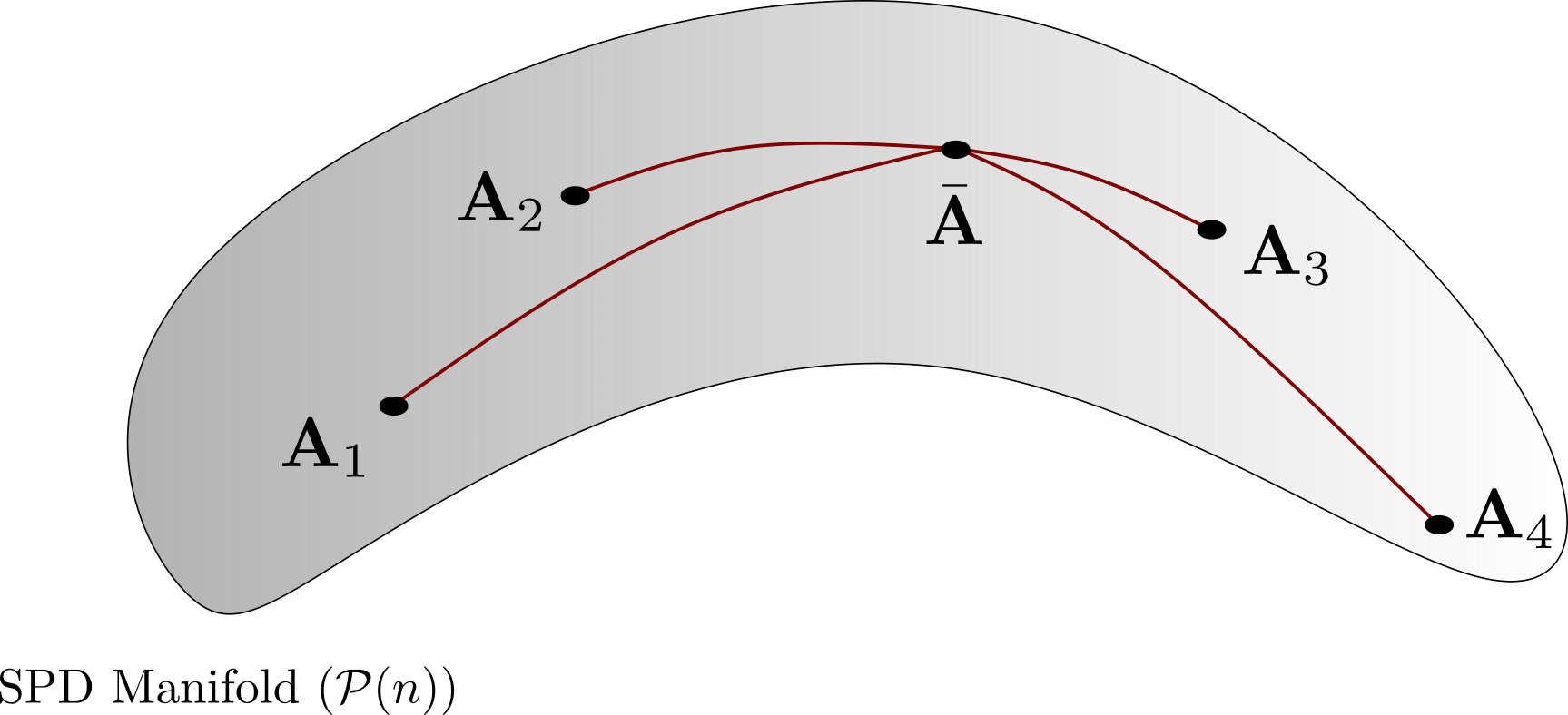}  
      \caption{Geometric mean.}
      \label{fig:sub-first}
    \end{subfigure}
    \begin{subfigure}{.49\textwidth}
      \centering
      \includegraphics[width=.9\linewidth]{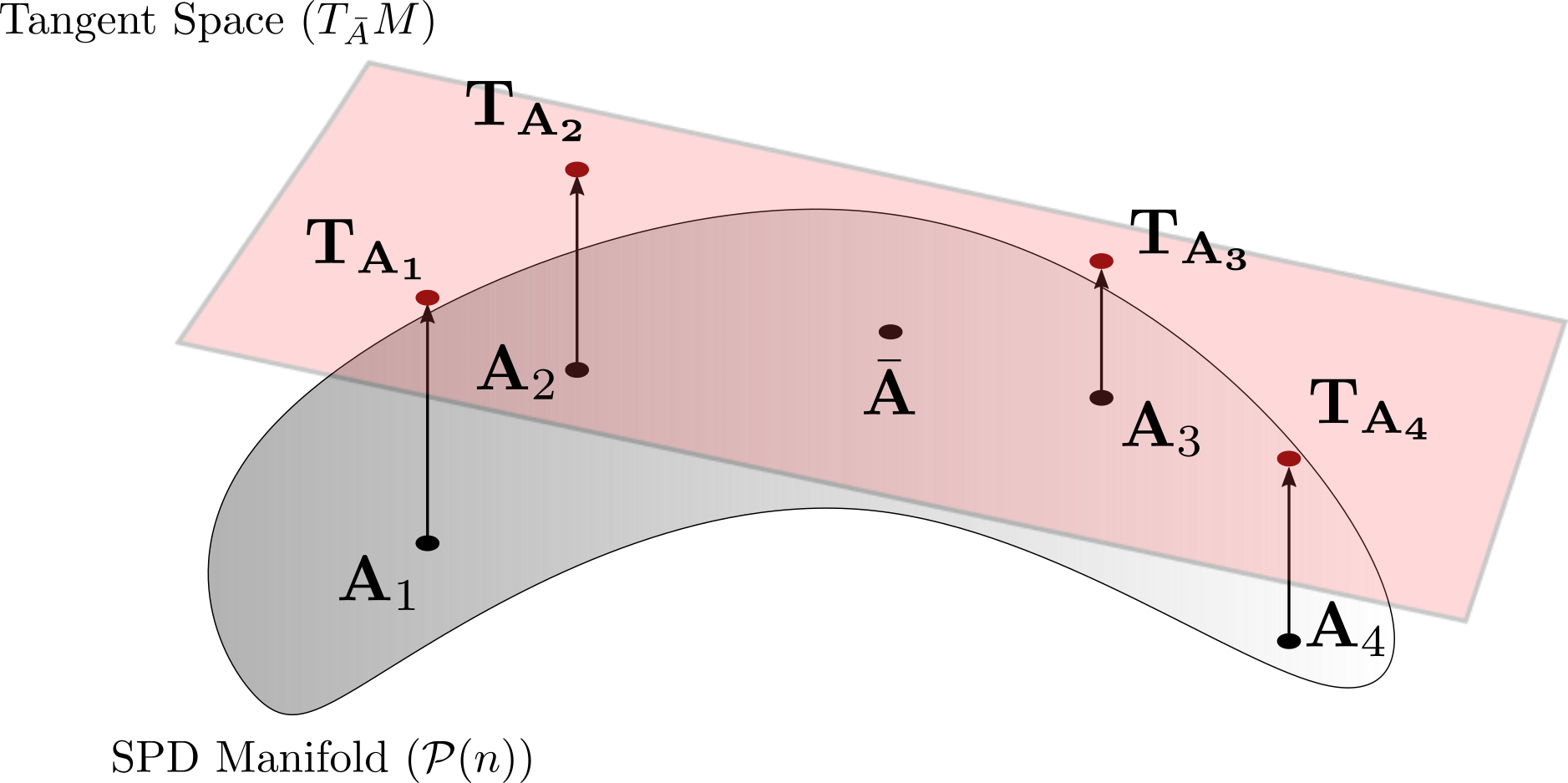}  
      \caption{Tangent space projection.}
      \label{fig:sub-second}
    \end{subfigure} 
  \caption{(a) Illustration of the geometric mean $\bar{\mathbf{A}}$ of a set of matrices $\{\mathbf{A}_1,\mathbf{A}_2,\mathbf{A}_3,\mathbf{A}_4\} \in \mathcal{P}(n)$. The geometric mean represents a point on the manifold that minimizes the geodesic to all other matrices in the set. (b) Representation of the tangent space $T_{\bar{A}}M$ at the geometric mean. The set of matrices $\mathbf{A}_i$ are projected (through the logarithmic map) onto this tangent space with minimal geometric distortion $\mathbf{T_{A_i}} \in T_{\bar{A}}M$.} 
  \label{fig:matrix_mean}
\end{figure}

In the analysis of a set of SPD matrices, we often need to identify a center point on the SPD manifold that will minimize the distortion of all geometric relationships between the matrices of the dataset when mapped to the tangent space. This matrix is the (Riemannian) geometric mean of the matrices \cite{you2021re}. For a set of SPD matrices $\mathbf{A}_i$, the geometric mean (see Figure \ref{fig:matrix_mean}) is the matrix $\bar{\mathbf{A}}$ that minimizes the sum of squared geodesic distances to all other matrices in the set:

\begin{equation}
\bar{\mathbf{A}} := \mathop{\text{argmin}}_{\mathbf{M}} \ \sum_{i=1}^n ||\text{log}(\mathbf{M}^{-1/2}\mathbf{A}_i\mathbf{M}^{-1/2})||_F
\end{equation}

We can see that the geometric mean is obtained by solving a matrix optimization problem. For the SPD manifold this problem is geodesically convex (similar to Euclidean convexity) \cite{absil2009optimization}. This optimization problem can be solved by using classical optimization algorithms that have been adapted to geometric setting (e.g., gradient descent) \cite{absil2009optimization}. A detailed review on these approaches can be found in the work by Absil, Mahony, and Sepulchre \cite{absil2009optimization}. 

Given a set of SPD matrices $\mathbf{A}_i$ and a geometric mean $\bar{\mathbf{A}}$, we can construct a tangent space at the geometric mean $T_{\bar{A}}M$, and project the SPD matrices onto the tangent space $\mathbf{T_{A_i}} = \text{log}_{\bar{A}}(\mathbf{A}_i)$, as shown in Figure \ref{fig:matrix_mean}. The matrices are now represented in a vector (linear) space that reflects the geometry of the SPD manifold. Projecting the data into a vector space allows us to apply common matrix analysis methods such as PCA.

\section{Case Study : Process Monitoring}

\begin{figure}[!h]
  \centering
  \includegraphics[width=.8\linewidth]{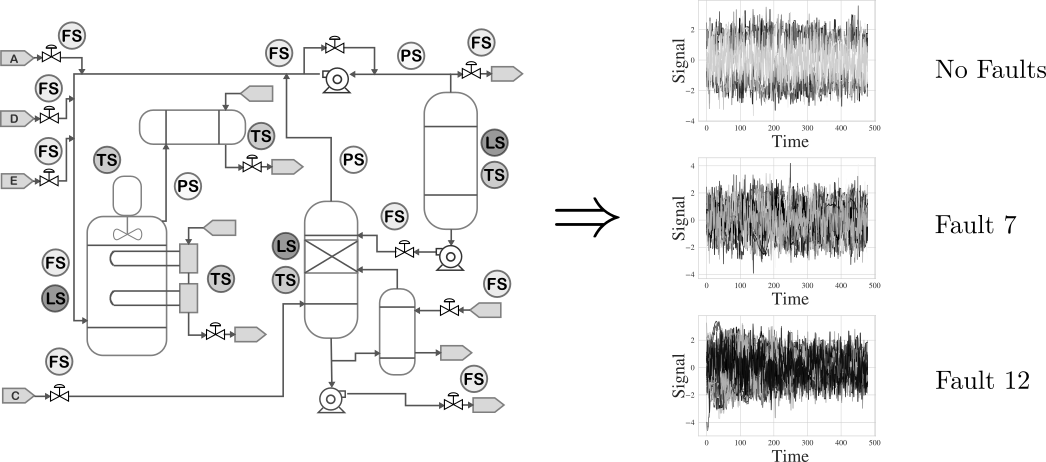} 
  \caption{Simplified illustration of the Tennessee Eastman Process (TEP) and resulting multivariate time series sensor data. Process sensors measure values such as temperature, pressure, flow, and level. From the multivariate time series data it is difficult to distinguish whether there is a fault occurring, or what type of fault may be occurring. Thus, a simplified and informative representation of the data is needed to be able to distinguish when the process is behaving normally or is experiencing a particular fault.}
  \label{fig:Ten_Est}
\end{figure}

We focus on data obtained from a simulated industrial process known as the Tennessee Eastman Process (TEP) \cite{downs1993plant}. This dataset is a widely used benchmark dataset for testing and comparing various anomaly (i.e. fault) detection methods \cite{ku1995disturbance,yin2012comparison,chiang2000fault}. Figure \ref{fig:Ten_Est} provides a high-level illustration of the process along with the multivariate time series data that is produced by the sensors monitoring the process. The process has a total of 52 measurements, 41 are process variables, 11 are manipulated variables. There are 20 different potential faults, which are defined in Table \ref{tab:case2fault} (for further details see Appendix A in  \cite{downs1993plant}).  Our aim is to use geometric methods to detect and classify the presence and type of fault using only the multivariate sensor data.  

\begin{table}[!h]
	\centering
	\begin{threeparttable}[!h]
		\caption{Types of Faults for Tennessee Eastman Process \cite{downs1993plant}.}
		\begin{tabular}{ccc}
			\toprule
			Fault ID   &                        Fault Name                        &       Type       \\ \midrule
			Fault 1   &    A/C feed ratio, B composition constant (stream 4)     &       Step       \\
			Fault 2   &       B composition, A/C ratio constant (stream 4)       &       Step       \\
			Fault 3   &              D feed temperature (stream 2)               &       Step       \\
			Fault 4   &         Reactor cooling water inlet temperature          &       Step       \\
			Fault 5   &        Condenser cooling water inlet temperature         &       Step       \\
			Fault 6   &                  A feed loss (stream 1)                  &       Step       \\
			Fault 7   & C header pressure loss - reduced availability (stream 4) &       Step       \\
			Fault 8   &           A, B, C feed composition (stream 4)            & Random variation \\
			Fault 9   &              D feed temperature (stream 2)               & Random variation \\
			Fault 10   &              C feed temperature (stream 4)               & Random variation \\
			Fault 11   &         Reactor cooling water inlet temperature          & Random variation \\
			Fault 12   &        Condenser cooling water inlet temperature         & Random variation \\
			Fault 13   &                    Reaction kinetics                     &    Slow drift    \\
			Fault 14   &               Reactor cooling water valve                &     Sticking     \\
			Fault 15   &              Condenser cooling water valve               &     Sticking     \\
			Fault 16-20 &                         Unknown                          &     Unknown      \\ \bottomrule
		\end{tabular}
		\label{tab:case2fault}
	\end{threeparttable}
\end{table}

The difficulty in distinguishing potential faults in the TEP is illustrated in Figure \ref{fig:Ten_Est}, where multivariate time series sensor signals for the 52 monitored variables are shown for the TEP without faults, and in the presence of fault 7 (step change in the component "C" header pressure) and fault 12 (random variation in the condenser cooling water inlet). The complexity of the multivariate process dynamics and of the number of sensor measurements make it difficult to reliably identify if a fault is occurring and to distinguish between fault types. Our method focuses on simplifying the data by quantifying the relationships between the 52 measured variables through covariance matrices and leveraging the geometry of if covariance matrices to detect and distinguish faults in the TEP. 

\subsection{Data Pre-Processing}

\begin{figure}[!h]
  \centering
    \begin{subfigure}{.3\textwidth}
      \centering
      \includegraphics[width=.95\linewidth]{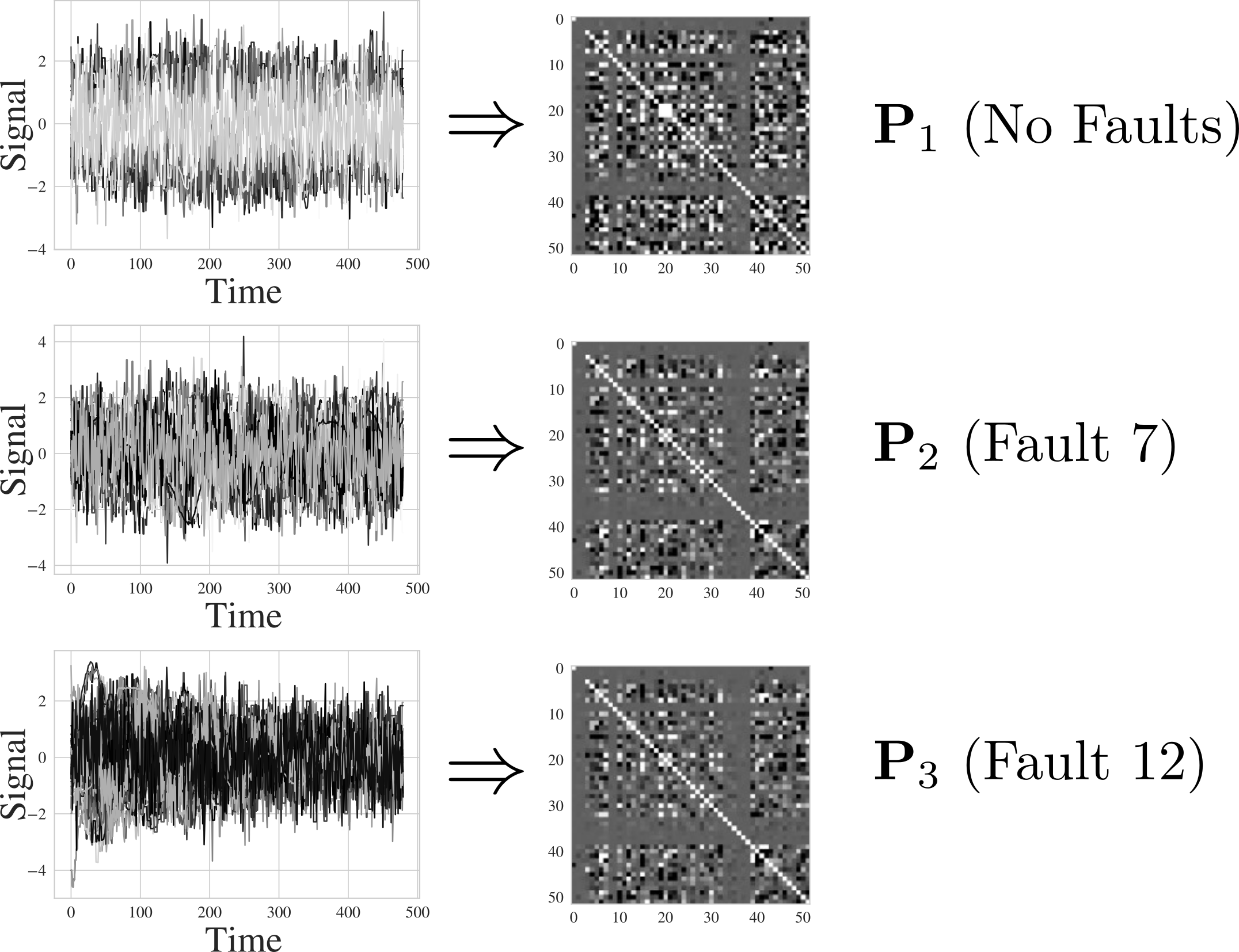}  
      \caption{Covariance matrices.}
      \label{fig:sub-first}
    \end{subfigure}
    \begin{subfigure}{.3\textwidth}
      \centering
      \includegraphics[width=.75\linewidth]{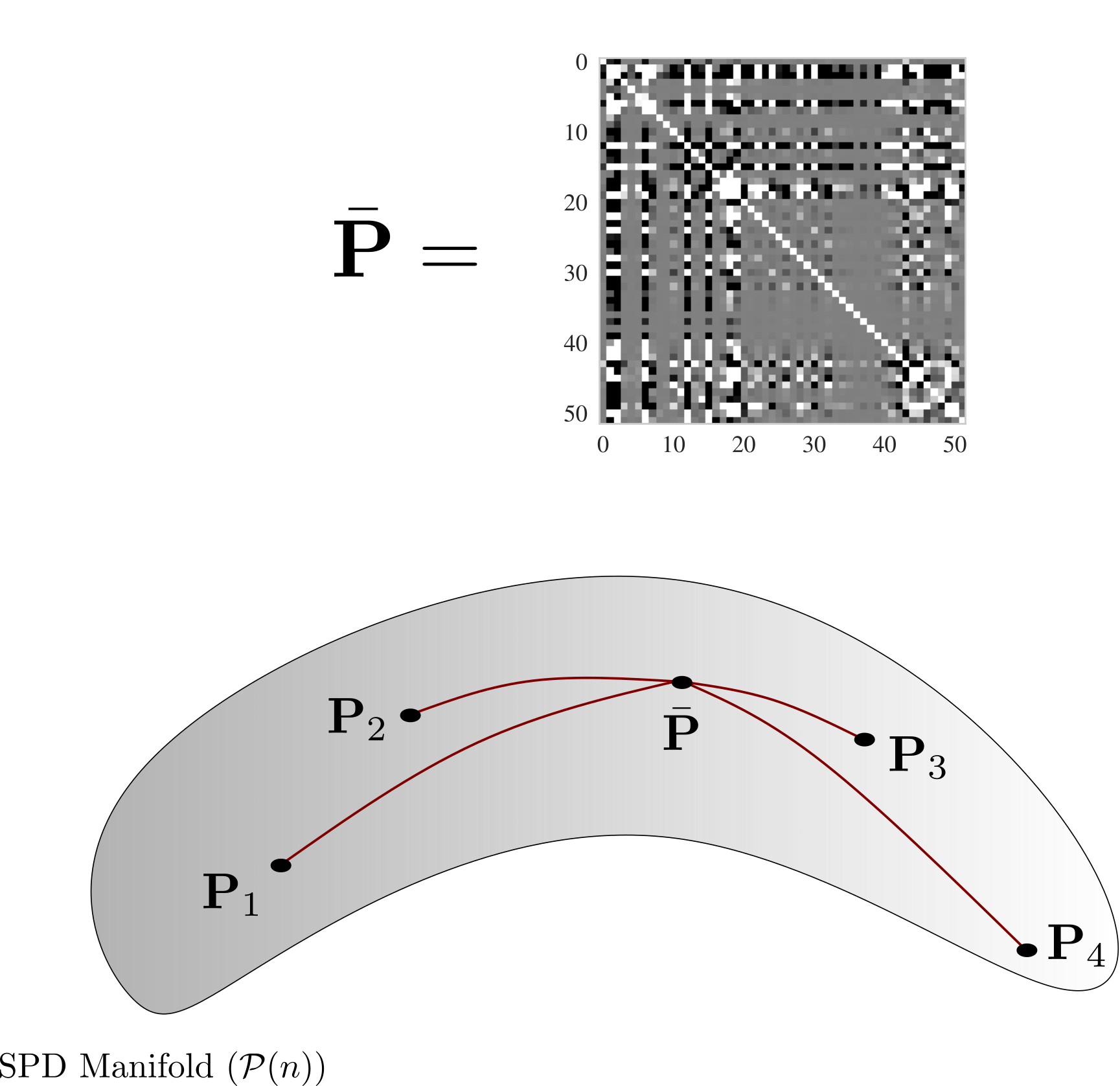}  
      \caption{Geometric mean.}
      \label{fig:sub-second}
    \end{subfigure} 
    \begin{subfigure}{.3\textwidth}
      \centering
      \includegraphics[width=.8\linewidth]{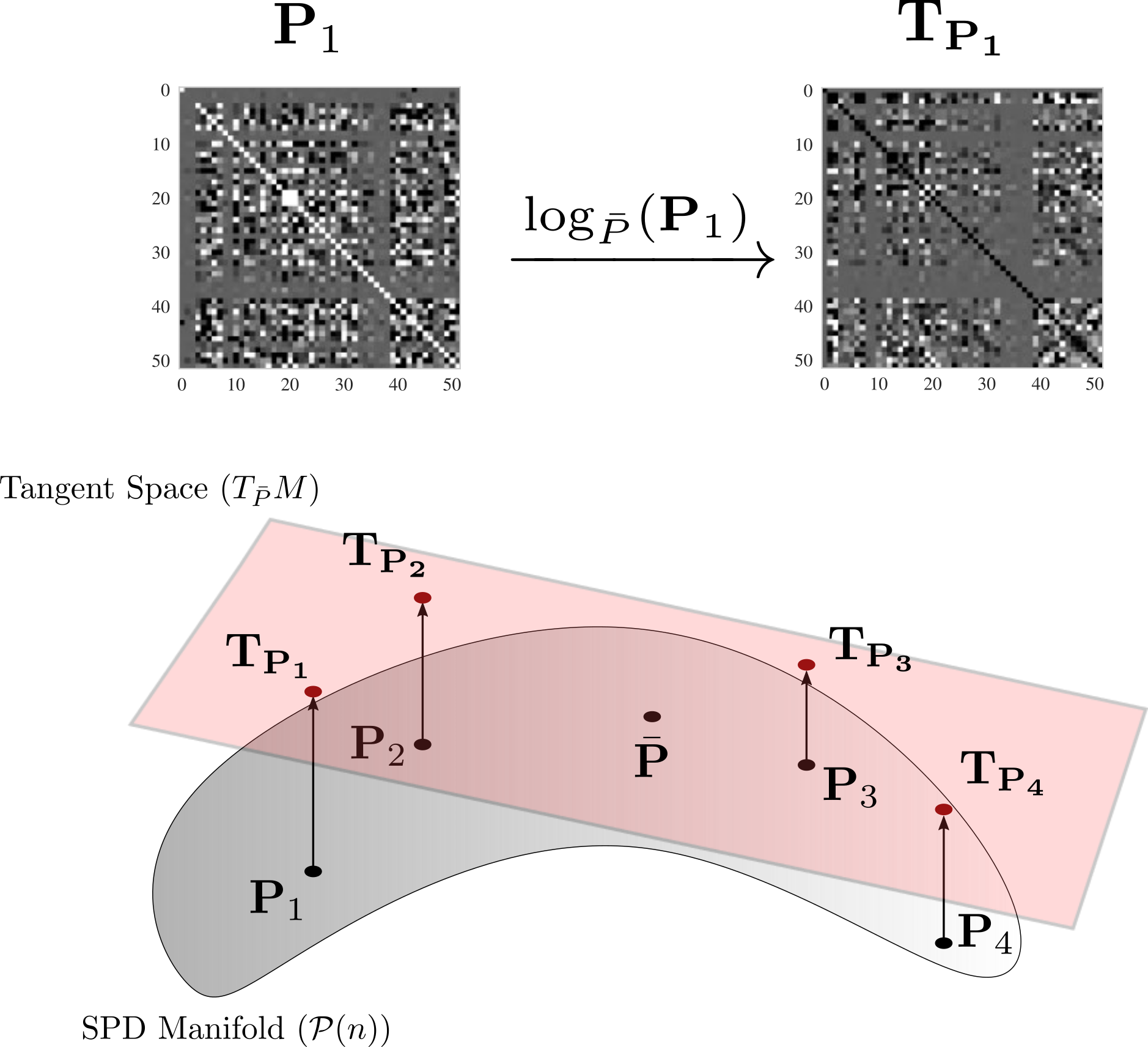}  
      \caption{Tangent space projection.}
      \label{fig:sub-second}
    \end{subfigure} 
  \caption{Representation of the data pre-processing workflow for the geometric analysis of the TEP data. (a) Covariance matrices are constructed from the sensor data multivariate time series forming a set of SPD matrices ($\mathbf{P}_i \in \mathcal{P}(n)$). (b) The geometric mean of the matrices ($\bar{\mathbf{P}} \in \mathcal{P}(n)$) is identified as the point that minimizes the squared geodesic distance to all other points. (c) All derived covariance matrices are mapped to the tangent space $T_{\bar{P}}M$ through the logarithmic mapping. This maps the matrices into a vector space that reflects the manifold geometry. The mapped data can then be analyzed with commonly defined dimensionality reduction and classification/regression methods.}
  \label{fig:data_workflow}
\end{figure}

To begin our analysis of the TEP, we must first pre-process the TEP data into multiple covariance matrices. To accomplish this, we represent each of the 52 measured variables as a univariate random variable $x_i$ where $i = 1,2,...,52$ and we denote the collection of signals as a multivariate random vector $\mathbf{X} = (x_1,...,x_n)$ where $n = 52$. We denote the observations of each signal at time $t = 1,2,..,m$ as $x_i(t) \in \mathbb{R}^m$.  We use this representation to construct the sample covariance matrix for the process data $\mathbf{P} \in \mathcal{P}(n)$ as:

\begin{align}
\mathbf{P} := \frac{1}{m-1}\mathbf{X}\mathbf{X}^T
\end{align}

The TEP dataset consists of multiple separate simulations of the process, both with and without faults. Thus, for each simulation we construct a sample covariance matrix $\mathbf{P}_i$. Our goal is to pair each simulation sample covariance matrix with the fault occurring in the simulation. Figure \ref{fig:data_workflow} provides examples of the sample covariance matrices that are constructed from simulations containing no faults, fault 7, and fault 12. We note that there are no obvious differences between the covariance matrices that would identify a given fault. These covariance matrices lie on the SPD manifold, and can be integrated into our geometric framework. An illustration of this computational workflow is found in Figure \ref{fig:data_workflow}. We compute the geometric mean $\bar{\mathbf{P}}$ for our set of covariance matrices $\mathbf{P}_i \in \mathcal{P}(n)$, the matrices are then projected to the tangent space $T_{\bar{P}}M$ centered at the geometric mean via the logarithmic mapping $\mathbf{T_{P_i}} = \text{log}_{\bar{P}}(\mathbf{P}_i)$. The data is now projected into a vector space that retains the geometric characteristics of the SPD matrices with minimal distortion, and can be integrated in dimensionality reduction and classification algorithms to perform analysis.

\subsection{Principal Geodesic Analysis}

\begin{figure}[!h]
  \centering
    \begin{subfigure}{.49\textwidth}
      \centering
      \includegraphics[width=.9\linewidth]{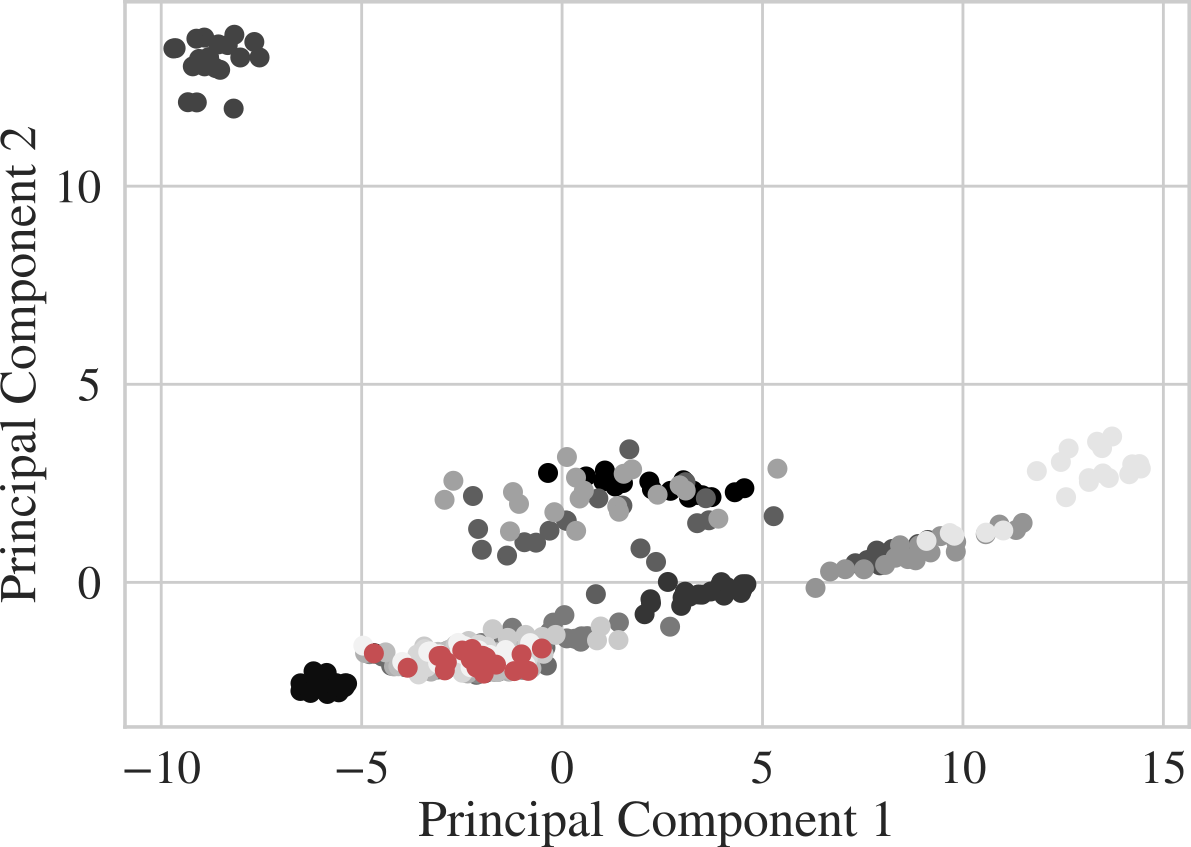}  
      \caption{Principal component analysis.}
      \label{fig:sub-first}
    \end{subfigure}
    \begin{subfigure}{.49\textwidth}
      \centering
      \includegraphics[width=.9\linewidth]{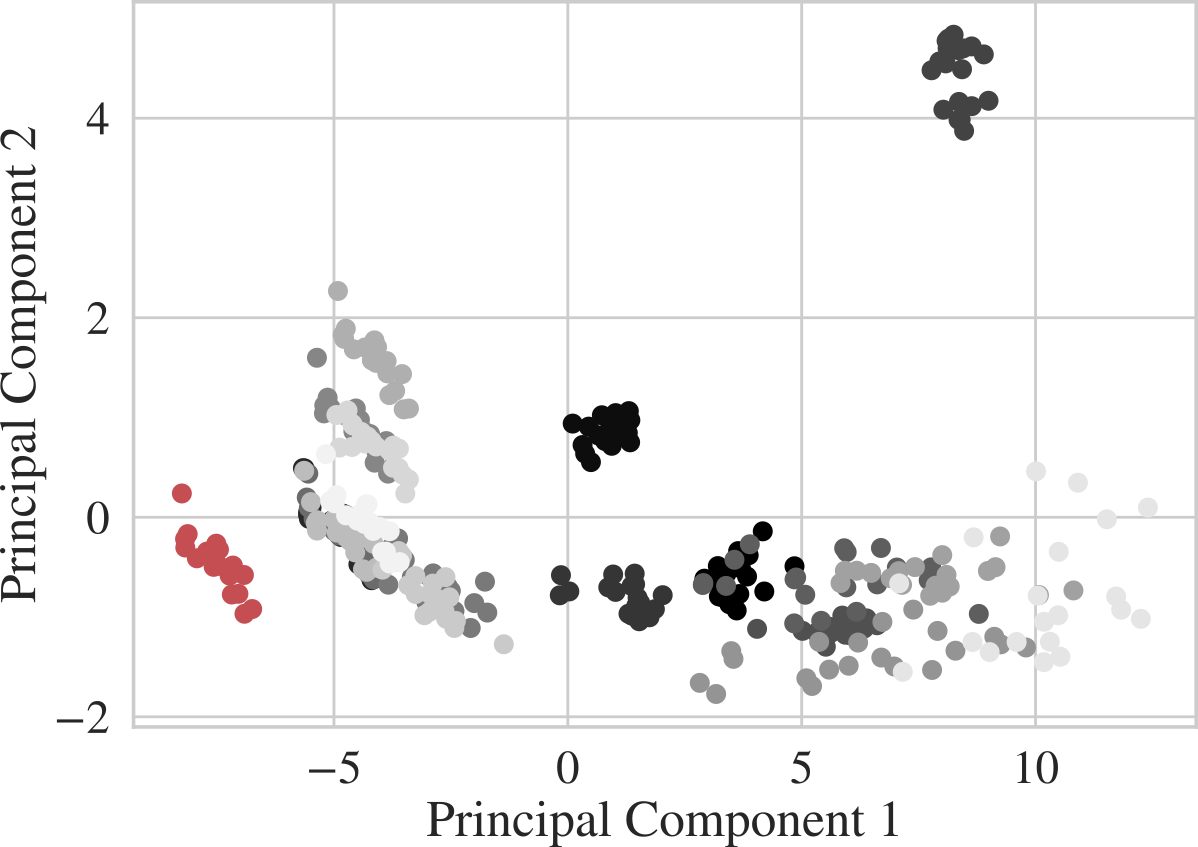}  
      \caption{Principal geodesic analysis.}
      \label{fig:sub-second}
    \end{subfigure} 
  \caption{Comparison of PCA applied to the raw covariance matrices and PCA applied to data mapped onto the tangent space (PGA). The red points represent the simulations where no fault is occurring and the grayscale points represent simulations with different faults. (a) PCA on the raw covariance matrices shows minimal separation in the data; faultless simulations are overlapped with faulty simulations. (b) PCA performed in the tangent space provides perfect separation between the faulty and faultless simulations, and also shows clustering of the faulty systems into separate groups. This demonstrates that simple considerations for the geometry of the data can yield improved results.}
  \label{fig:PCA_comp}
\end{figure}

Mapping the process data covariance matrices to the tangent space provides an avenue for the application of common dimensionality reduction techniques. Here, we apply PCA to the matrices mapped to the tangent  (vector) space. PCA applied on the tangent space of the SPD manifold is commonly known as Principal Geodesic Analysis (PGA), as it identifies the geodesics that capture the most variance in the data \cite{harandi2014manifold}. An example comparison of PGA versus PCA (directly on the covariance matrices) is presented in Figure \ref{fig:PCA_comp}. The simulations with no faults are colored in red and the faulty simulations are represented by different grayscale values. We can see that using only the first two components in PGA, we are able to perfectly separate the faulty and non-faulty simulations; on the other hand, when applying PCA directly on the covariance matrices we can see that there is significant overlap between the faulty and non-faulty simulations. The comparison of these projections demonstrates that capturing the geometry of the Riemannian manifold in the analysis of covariance matrices can improve the performance with minimal added complexity. 

This improvement in separation of the data through the geometric approach is due in part to the congruence invariance of our defined metric on the Riemannian manifold. As previously stated, congruence invariance means that any $n \times n$ invertable matrix $\mathbf{X}$ applied to a set of covariance matrices $\mathbf{P}_i \in \mathcal{P}(n)$ does not impact the geodesic distance between the two matrices:

\begin{align}
d_g(\mathbf{X}^T \mathbf{P}_i \mathbf{X}, \mathbf{X}^T \mathbf{P}_j \mathbf{X}) = d_g(\mathbf{P}_i,\mathbf{P}_j)
\end{align}

Operations such as re-scaling and normalization, which can be represented algebraically as invertable square matrices, have no impact on the geodesic distance between the matrices \cite{barachant2014plug}. Therefore, there is no need to select specific scaling or normalization strategies when applying our geometry based analysis of the data (e.g., PGA). However, this is not true when ignoring the data geometry, making methods such as PCA susceptible to the chosen framework (or lack of) for normalization/scaling. We perform no scaling prior to PCA in this case to ensure a direct comparison between PCA and PGA.

\subsection{Classification and Clustering Results}

\begin{figure}[!h]
  \centering
    \begin{subfigure}{.49\textwidth}
      \centering
      \includegraphics[width=.9\linewidth]{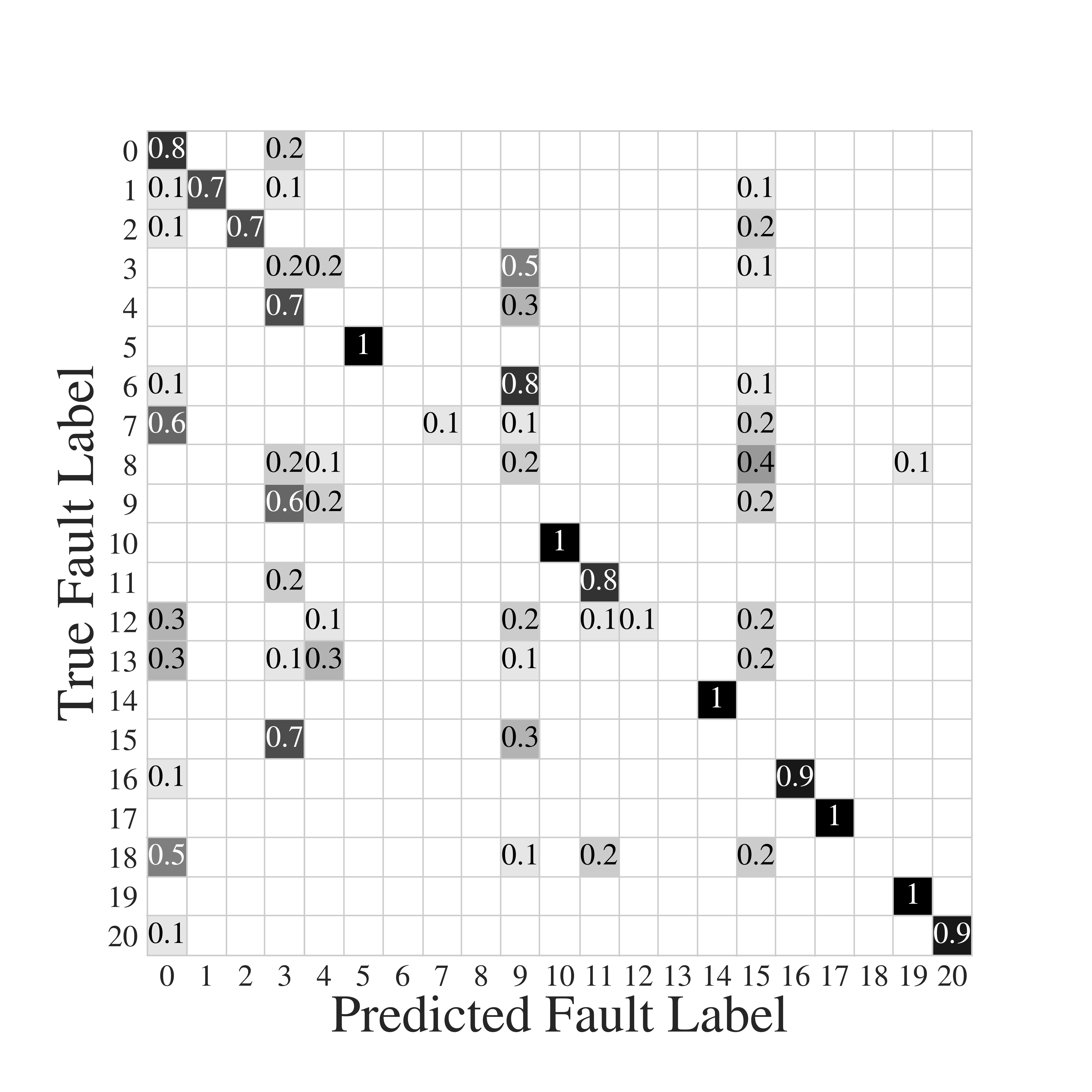}  
      \caption{Classification raw covariance matrices.}
      \label{fig:sub-first}
    \end{subfigure}
    \begin{subfigure}{.49\textwidth}
      \centering
      \includegraphics[width=.9\linewidth]{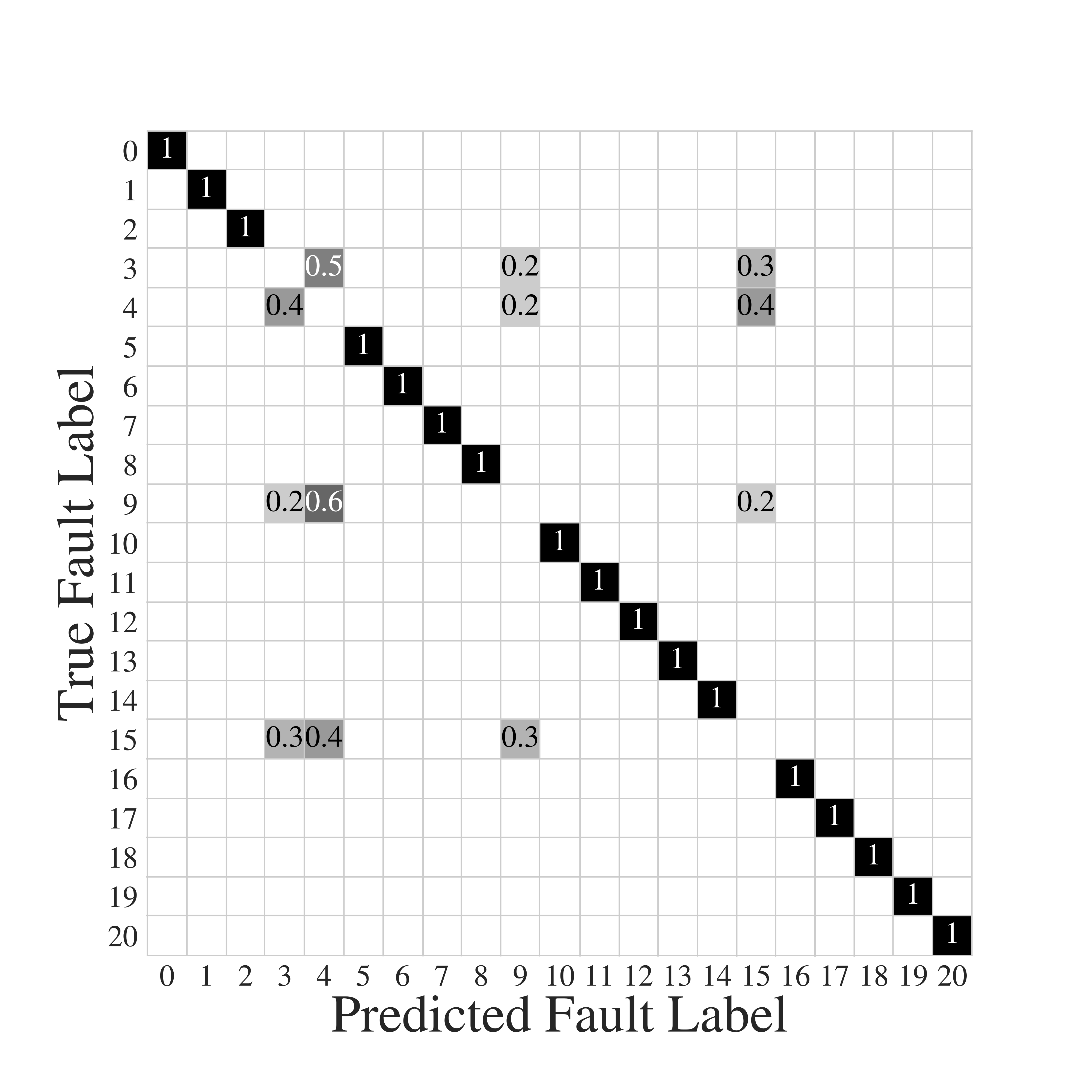}  
      \caption{Classification in tangent space ($T_{\bar{P}}M$).}
      \label{fig:sub-second}
    \end{subfigure} 
  \caption{Comparison of linear classification on the raw covariance matrices versus matrices mapped to the tangent space $T_{\bar{P}}M$. (a) Classification of the covariance matrices without regard for the data geometry results in poor classification accuracy. (b) Simple mapping of the data to the appropriate tangent space provides a dramatic improvement in classification accuracy.}
  \label{fig:TEP_Class}
\end{figure}

PGA analysis of the covariance matrices also reveals that there is definite clustering of the data with respect to the different fault types within the TEP dataset. This suggests that classification of the different fault types can be done directly using the tangent space of the SPD manifold. To investigate this we use a simple linear (ridge) classifier. We compare the prediction accuracy of the linear classifier using coefficients of the tangent space projected matrices versus the coefficients of the non-transformed covariance matrices as input. In the analysis, we perform a simple train-test split of the data, where $30\%$ of the data is used for testing and $70\%$ of the data is use for training. Figure \ref{fig:TEP_Class} illustrates the dramatic increase in accuracy when the model incorporates geometric information, which is reflected in the normalized confusion matrices. Here, a value $x \in [0,1]$ on the diagonal indicates an accuracy of $x*100\%$ when classifying a particular fault. All values in the off diagonal (e.g., row $i$, column $j$, where $i \neq j$) represent the percentage of covariance matrices associated with fault $i$ that have been incorrectly labeled as experiencing fault $j$. When the SPD manifold is accounted for via the tangent space projection, there is perfect classification of the data (with the exception of faults $3,4,9$ and $15$). When the manifold geometry is ignored, there are few instances where high classification accuracy is achieved. The faults $3,4,9$ and $15$ have been shown in prior work to be difficult to classify \cite{yin2012comparison}. We also note that these faults are only misclassified within their group (are never classified as having no fault), which suggests that there is limited quantifiable difference in the covariance matrices for these faults. The inclusion of more information around these particular faults may correct this issue and further increase accuracy.

\section{Case Study - Image Analysis}

\begin{figure}[!h]
  \centering
    \begin{subfigure}{.49\textwidth}
      \centering
      \includegraphics[width=.7\linewidth]{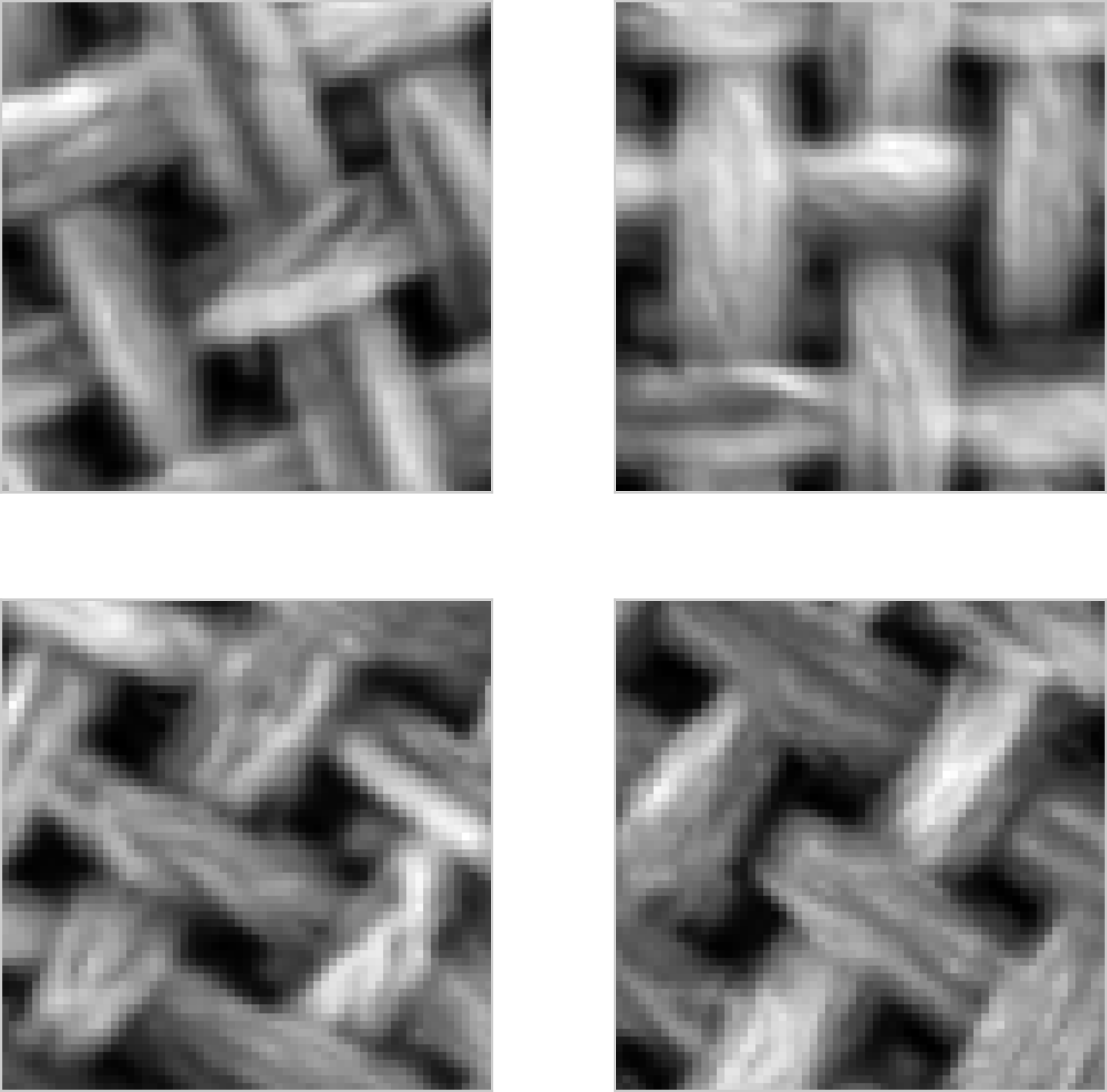}  
      \caption{Normal textiles.}
      \label{fig:sub-first}
    \end{subfigure}
    \begin{subfigure}{.49\textwidth}
      \centering
      \includegraphics[width=.7\linewidth]{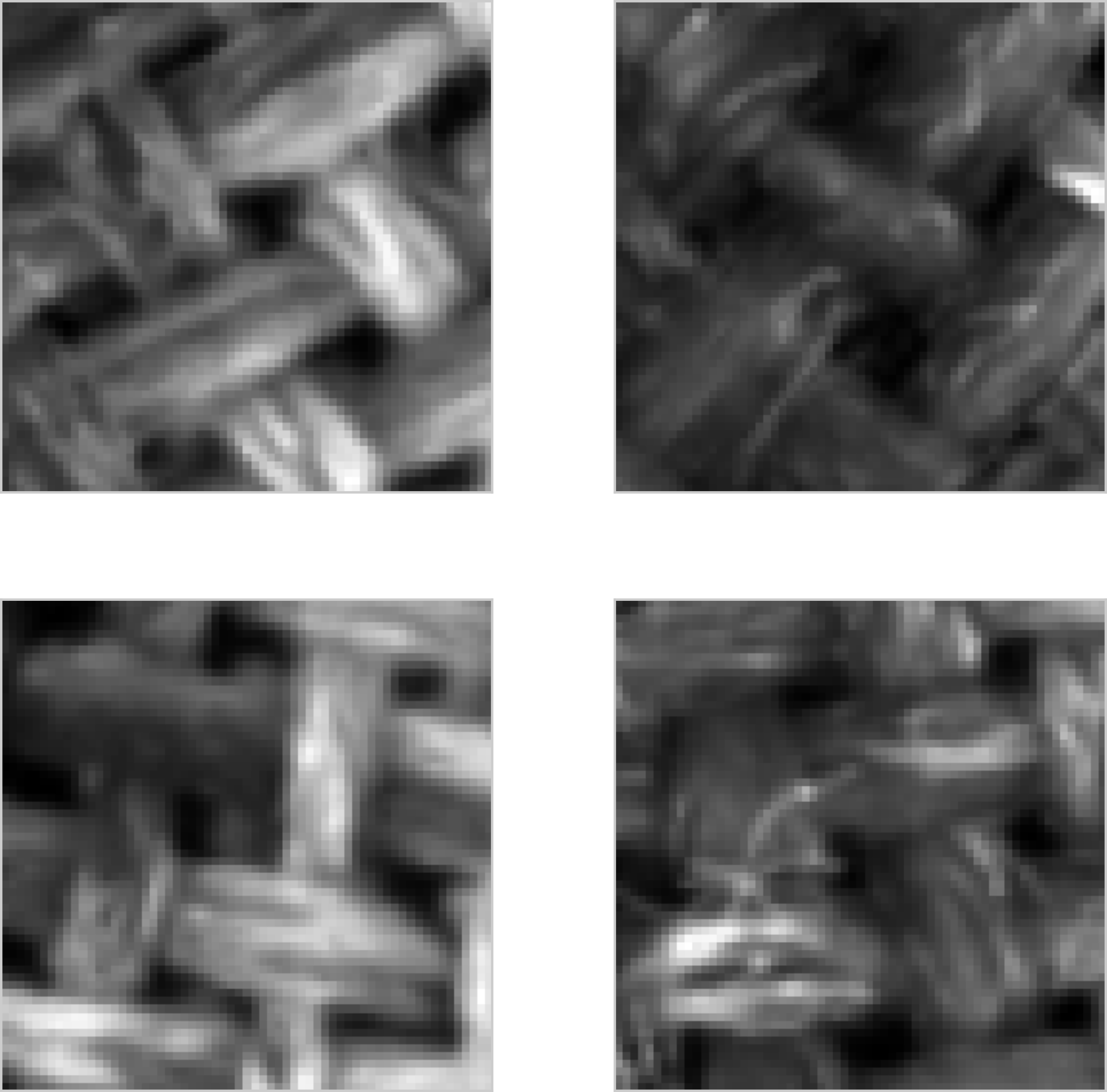}  
      \caption{Defective textiles.}
      \label{fig:sub-second}
    \end{subfigure} 
  \caption{Example images from the woven textile dataset \cite{bergmann2019mvtec}. (a) Representative sample of woven textiles with no defects. (b) Representative sample of woven textiles that are considered defective.}
  \label{fig:textile}
\end{figure}

Another important application of covariance matrices is in the analysis of images \cite{porikli2006fast,dryden2009non,caefer2008improved}. Here, we focus on an analysis of real images from textile manufacturing that contain non-defective and defective woven textiles taken. The images were obtained from the public MVTEC AD dataset \cite{bergmann2019mvtec}. Example images of non-defective and defective textiles are found in Figure \ref{fig:textile}. Our goal is to classify textiles using a linear classifier and Riemannian geometry.  

\subsection{Data Pre-Processing}

\begin{figure}[!h]
  \centering
  \includegraphics[width=.8\linewidth]{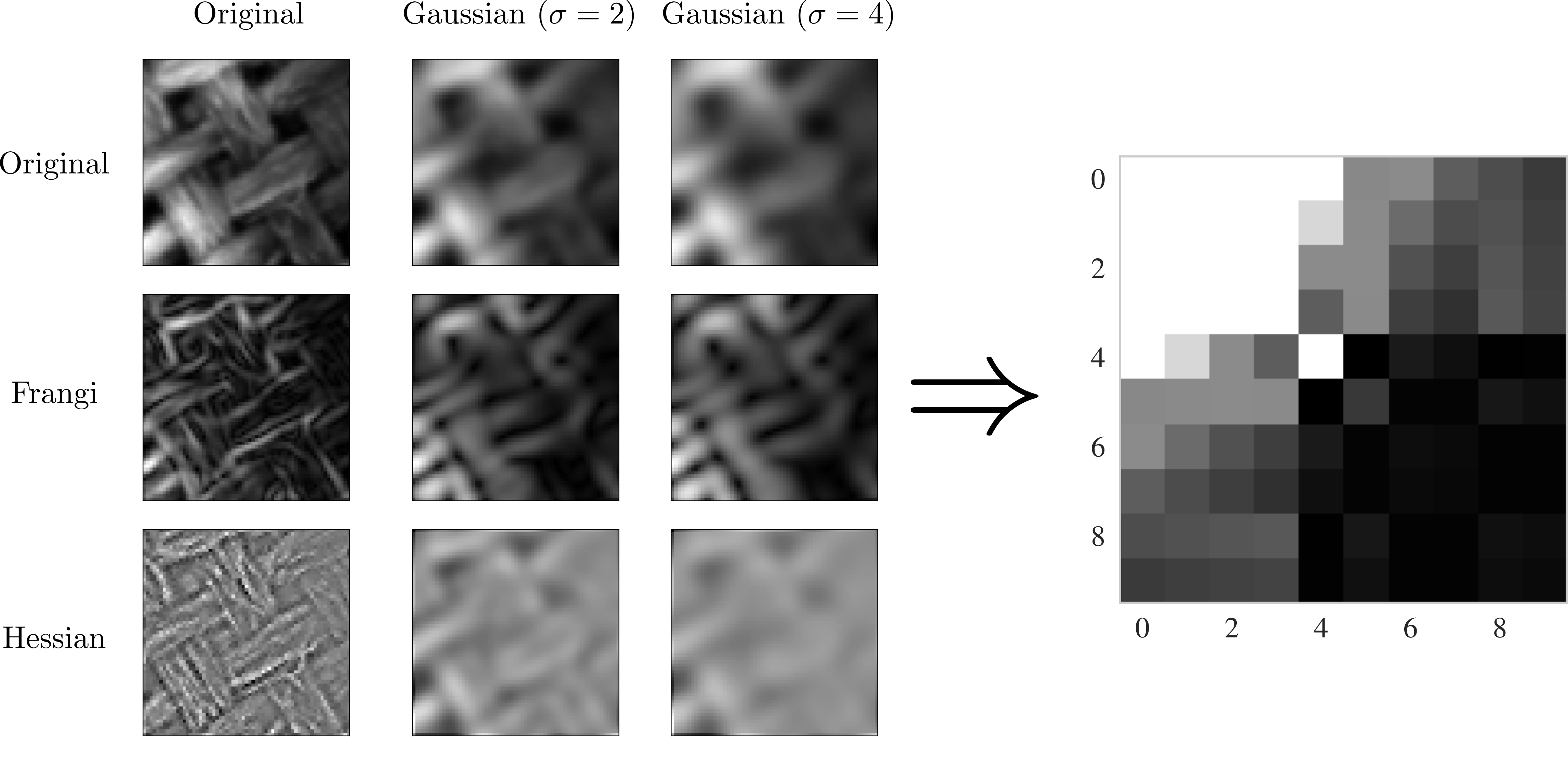} 
  \caption{Workflow for the construction of an image covariance matrix. (left) Image filters and transformations emphasize specific characteristics of an image. Gaussian filters emphasize features of different scale within an image, and the Frangi and Hessian filters capture important fiber and edge features of an image. (right) The covariance between each image representation can be computed and used to form a covariance matrix. Importantly, the covariance matrix representation is invariant to transformations such as rotation and translation which are present in the textile images.}
  \label{fig:filters}
\end{figure}

Covariance matrices are useful data representations for images because they are invariant to translations and rotations \cite{tuzel2006region}. These matrices can also be used to combine multiple image features that can be quantified through filters and kernel methods \cite{tuzel2008pedestrian}. Here, we select nine image features, as shown in Figure \ref{fig:filters}. We implement both the Frangi and Hessian filters, which are designed to detect edges and fiber structures of the woven textures \cite{frangi1998multiscale}. We apply these filters to the image; multiple transformed images that are smoothed through a Gaussian filter of varying strength are shown in Figure \ref{fig:filters}. We use Gaussian filters to emphasize features of different scale within the image \cite{young1995recursive}. This yields nine total feature images (including the original image), which we use to construct a sample covariance matrix. We do this by taking each feature image (which we treat as a $64 \times 64$ matrix) and vectorizing each image $x_i \in \mathbb{R}^{m}$, where $i = 1,2,...,9$ and $m = 4096$. Each feature image can therefore be represented as a realization of a multivariate random vector $\mathbf{X} = (x_1,x_2,...,x_9)$. We use this observation to construct a sample covariance matrix $\mathbf{P} \in \mathcal{P}(n)$ (see Figure \ref{fig:filters}) where $n=9$:

\begin{equation}
\mathbf{P} := \frac{1}{m-1}\mathbf{X}\mathbf{X}^T.
\end{equation}

The covariance matrices are SPD, and can be directly integrated into our geometric analysis framework.

\subsection{Classification Results}

\begin{figure}[!h]
  \centering
    \begin{subfigure}{.49\textwidth}
      \centering
      \includegraphics[width=1\linewidth]{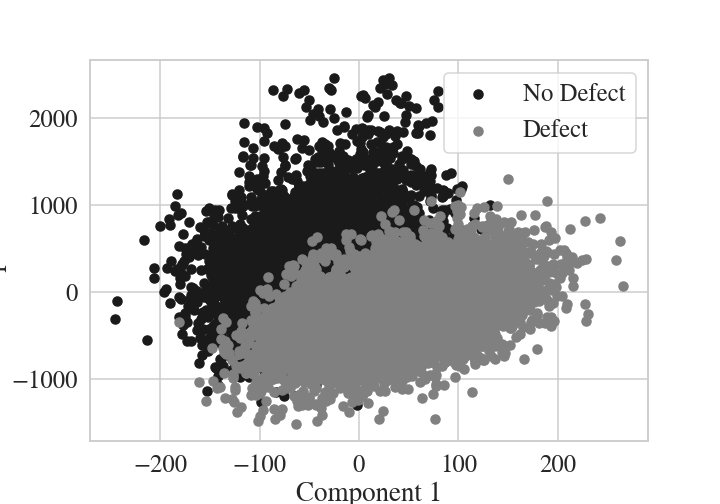}  
      \caption{PGA on covariance matrices.}
      \label{fig:sub-first}
    \end{subfigure}
    \begin{subfigure}{.49\textwidth}
      \centering
      \includegraphics[width=1\linewidth]{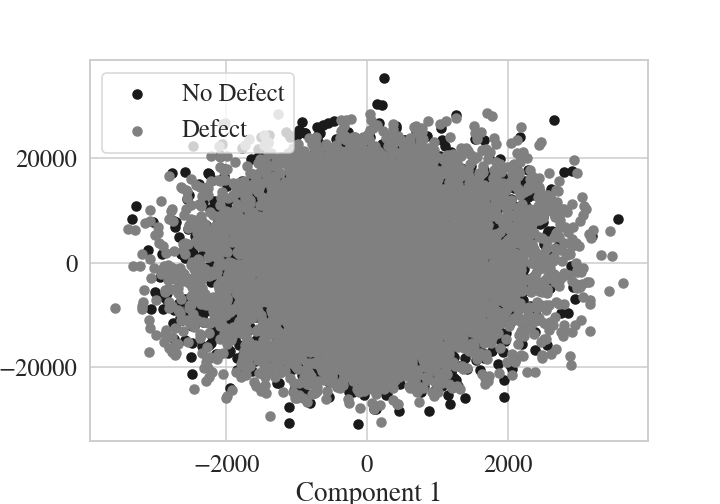}  
      \caption{PCA on raw images.}
      \label{fig:sub-second}
    \end{subfigure} 
  \caption{(a) PGA of the covariance matrices derived from the set of woven textile images. The analysis reveals clustering of the data into groups representing the defective and non-defective textiles. A simple linear SVM  classifier is able to separate the defective and non-defective textile images with $92 \%$ accuracy. (b) PCA analysis of the raw images; without the covariance matrix representation and considerations for the data geometry, we see there is almost no separation in the data.}
  \label{fig:textile_pca}
\end{figure}

We first apply PGA to the image covariance matrices (see Figure \ref{fig:textile_pca}). PGA reveals a distinct clustering and separation of the data with only two components. The PGA suggests that a simple classifier can be used to separate the defective and non-defective textile samples (after projecting to the tangent space). Thus, we construct a simple linear support vector machine classifier, which we train with $70\%$ of the image data and test our trained model on the remaining $30\%$ of the data. The covariance representation of the images provides a simple characterization of the data and its various features, while also imbuing the data with the inherent geometry of SPD matrices. This results in the trained model being able to separate the defective and non-defective samples in the testing data with $92 \%$ accuracy. We also compare our PGA analysis of the covariance matrix representation to PCA applied to the raw image data in Figure \ref{fig:textile_pca}. The PCA analysis reveals almost no separation of the data, and results in poor classification accuracy when used as input to a linear SVM (50 \% accuracy). This is likely due to sensitivity of the analysis on the raw images to rotations and translations of the textiles.

\section{Conclusions and Future work}

We have presented an introduction to concepts of Riemannian geometry for symmetric positive definite (SPD) matrices and show how these concepts can be used in applications of interest to chemical engineers. Specifically, we discussed approaches to capture the geometry of the SPD manifold (a Riemannian manifold) in dimensionality reduction and classification tasks. Through a couple of case studies, we demonstrated that capturing such geometry can lead to significant improvements in accuracy and interpretability.  This work focused primarily on the manifold of SPD matrices, but we note that similar concepts can be applied to over matrix representations such as fixed-rank matrices, orthogonal matrices, and shape spaces as well as to other datasets where a Riemannian manifold is known to exist (e.g., points on a sphere). Furthermore, geometry can be leveraged in optimization and control to relax certain constraints that are already enforced by the underlying system geometry (e.g., optimization on the SPD manifold ensures that all solutions are SPD). 

\section*{Acknowledgments} 

We acknowledge funding from the U.S. National Science Foundation (NSF) under BIGDATA grant
IIS-1837812. 

\bibliography{References.bib}

\end{document}